\documentclass[a4paper,11pt]{article}
\pdfoutput=1 

\usepackage{jheppub} 
\usepackage[T1]{fontenc} 

\usepackage{orcidlink}
\usepackage{xcolor}
\usepackage{xfrac}
\usepackage{tabularray}
\usepackage[frozencache,cachedir=minted-cache]{minted}
\UseTblrLibrary{booktabs}
\UseTblrLibrary{siunitx}
\newcommand{\bftab}{\fontseries{b}\selectfont} 
\usepackage{mathtools}
\usepackage{csquotes}
\usepackage{multirow}
\usepackage{microtype}
\usepackage{tikz}
\usetikzlibrary{cd}
\graphicspath{{figures/}}

\newcommand{\acronymsw}[1]{\textsc{#1}}
\newcommand{\Pepper}{\acronymsw{Pepper}}
\newcommand{\Chili}{\acronymsw{Chili}}
\newcommand{\Comix}{\acronymsw{Comix}}
\newcommand{\Vegas}{\acronymsw{Vegas}}
\newcommand{\Sherpa}{\acronymsw{Sherpa}}
\newcommand{\NNPDF}{\acronymsw{NNPDF}}
\newcommand{\LHAPDF}{\acronymsw{LHAPDF}}
\newcommand{\HighFive}{\acronymsw{HighFive}}
\newcommand{\acronym}[1]{#1}
\newcommand{\LHC}{\acronym{LHC}}

\newcommand{\dif}{\mathop{}\!\mathrm{d}}

\newcommand\pd[3][]{\frac{\partial\spx{#1}#2}{\partial#3\spx{#1}}}
\newcommand{\od}[3][]{\frac{\dif\spx{#1}#2}{\dif#3\spx{#1}}}

\newcommand{\symscr}[1]{\mathcal{#1}}

\newcommand{\real}{\mathbb{R}}
\DeclarePairedDelimiter\abs{\lvert}{\rvert}

\DeclarePairedDelimiter\intervalrightopen{[}{)}

\DeclarePairedDelimiter{\mean}{\langle}{\rangle}%

\DeclareMathOperator{\var}{Var}

\makeatletter
\newcommand{\spx}[1]{%
  \if\relax\detokenize{#1}\relax
    \expandafter\@gobble
  \else
    \expandafter\@firstofone
  \fi
  {^{#1}}%
}
\makeatother

\hypersetup{pdfauthor={Enrico Bothmann, Timo Janßen, Max Knobbe, Bernhard Schmitzer, Fabian Sinz},
  pdftitle={Efficient many-jet event generation with Flow Matching},
  pdfkeywords={QCD, Monte Carlo, Flow Matching}
}

\definecolor{listingsbg}{rgb}{0.95,0.95,0.95}

\title{\boldmath Efficient many-jet event generation with Flow Matching}

\author[a,c]{E.\ Bothmann \orcidlink{0000-0001-6786-6843},}
\author[b,c]{T.\ Janßen \orcidlink{0000-0001-9466-477X},}
\author[d]{M.\ Knobbe \orcidlink{0000-0002-6632-6869},}
\author[b,e]{B.\ Schmitzer}
\author[b,e]{and F.\ Sinz}

\affiliation[a]{IT Department, CERN, 1211 Geneva 23, Switzerland}
\affiliation[b]{Campus Institute Data Science, University of Göttingen, Germany}
\affiliation[c]{Institute for Theoretical Physics, University of Göttingen, Germany}
\affiliation[d]{Theoretical Physics Division, Fermi National Accelerator Laboratory, USA}
\affiliation[e]{Institute for Computer Science \& Campus Institute Data Science, University of Göttingen, Germany}

\emailAdd{enrico.bothmann@cern.ch}
\emailAdd{tjansse2@uni-goettingen.de}
\emailAdd{mknobbe@fnal.gov}
\emailAdd{schmitzer@cs.uni-goettingen.de}
\emailAdd{sinz@uni-goettingen.de}

\abstract{%
  We apply for the first time the Flow Matching method
  to the problem of phase-space sampling for event generation
  in high-energy collider physics.
  By training the model
  to remap the random numbers
  used to generate the momenta and helicities
  of the scattering matrix elements
  as implemented in the portable partonic event generator \Pepper,
  we find substantial efficiency improvements in the studied processes.
  We focus our study on the highest final-state multiplicities in Drell--Yan and top--antitop pair production
  used in simulated samples for the Large Hadron Collider, which computationally are the most relevant ones. We find that the
  unweighting efficiencies improve by factors of 150 and 17, respectively,
  when compared to the standard approach
  of using a \Vegas-based optimisation.
  We also compare Continuous Normalizing Flows trained with Flow Matching
  against the previously studied Normalizing Flows based on Coupling Layers
  and find that the former leads to better results, faster training
  and a better scaling behaviour across the studied multiplicity range.
}

\begin{document}

\maketitle
\flushbottom

\section{Introduction}
\label{sec:intro}
The excellent experimental precision of the current and future runs
at the Large Hadron Collider (LHC) and its successors
must be met with a corresponding increase in the precision of theoretical simulations.
This is essential to meet the objectives outlined in collider physics programs.
Otherwise the uncertainty of many experimental results would be dominated
by the uncertainty of the simulated data samples
that enter experimental analyses in many ways,
most prominently to provide background estimates~\cite{EuropeanStrategyGroup:2020pow,Narain:2022qud,HEPSoftwareFoundation:2017ggl,HSFPhysicsEventGeneratorWG:2020gxw,%
  HSFPhysicsEventGeneratorWG:2021xti}.
Precision must be improved across three key areas: parametric accuracy, numerical stability, and statistical precision.
In this contribution, we focus on the latter,
i.e.\ the challenge to deliver large simulated data samples
that match the statistical quality of the experimental data.
To illustrate the scale of this challenge, the ATLAS collaboration estimates~\cite{ATLAS:2021yza} that
330 billion simulated events are required to match the HL-LHC dataset for a key background process
vector boson production in association with additional jets,
when using the general-purpose event generator \Sherpa\ v2.2.11.
This translates to 3.8~million HEPSPEC06 years~\cite{ATLAS:2021yza},
which corresponds to about
1000 Perlmutter 2$\times$CPU nodes for an entire year~\cite{ATLAS:2021yza}.

At the heart of the event-generation pipeline lies the hard scattering event,
which has two incoming particles for collider setups, 
and a fixed number
of outgoing particles---up to seven in current production campaigns.
Except for the experiment-specific simulation of the detector response,
it is by far the most computationally demanding part of the pipeline,
and therefore a prime target for performance improvements~\cite{Bothmann:2022thx}.
Low overall Monte-Carlo simulation efficiency is driven by the highest-multiplicity processes in the simulation,
for which current simulation programs
only achieve very low phase-space sampling efficiency.
They typically rely  on multi-channel techniques~\cite{Kleiss:1994qy} combined with the adaptive sampling
algorithm \Vegas\ (or similar \emph{classical} Machine Learning algorithms)~\cite{Lepage:1977sw,Ohl:1998jn,Lepage:2020tgj,
  Jadach:2002kn,Hahn:2004fe,vanHameren:2007pt}.
However, for processes with 7 outgoing particles, corresponding to a 19-dimensional phase space,
the overall efficiency achieved with these methods can be as low as 0.01\,\%~\cite{Hoche:2019flt},
thus immensely degrading the statistical power of the generated sample.

Modern Machine Learning approaches to optimise sampling based on deep neural networks
have been proposed to improve the efficiency by making use of their more flexible parametrisation.
There has been a particular focus on using Normalizing Flows as a drop-in replacement for \Vegas~\cite{Klimek:2018mza,Bothmann:2020ywa,
 Gao:2020zvv,Heimel:2022wyj,Verheyen:2022tov,Heimel:2023ngj,Heimel:2024wph,Butter:2022rso}.
For individual partonic channels, efficiency improvements of factors between 5 and 10 over \Vegas\
have been demonstrated for up to five final-state particles~\cite{Heimel:2023ngj}.
However, a systematic study which addresses the highest and most expensive multiplicities
used in standard productions at the LHC is still missing.

Here, we propose for the first time to solve this issue using Continuous Normalizing Flows~\cite{Chen:2018}, which implement time-continuous flows through an ordinary differential equation (ODE). To avoid expensive simulation of the ODE dynamics during training, we use the Flow Matching method~\cite{Lipman:2023,Albergo:2023building,Albergo:2023stochastic,Liu:2022}.%
\footnote{%
Flow Matching has recently been studied
for neural-network based generation of LHC events and off-shell effects~\cite{Butter:2023fov,Favaro:2025psi}.
In contrast, in our study we do not model physical effects with a generative model,
but use Flow Matching only to improve the simulation's sampling efficiency.} 
We study the performance of our approach for the computationally by far most prominent simulated processes for the LHC,
i.e.\ lepton pair production ($pp \to e^+e^-$ + up to 5 jets) and top quark pair production ($pp \to t \bar t$ + up to 4 jets),
utilising the fast event generation framework \Pepper~\cite{Bothmann:2023gew}
and its phase-space sampler \Chili~\cite{Bothmann:2023siu}.
This covers the entire multiplicity range of current production campaigns.
We evaluate our results in several key measures,
e.g.\ in the integration error of the generated sample,
the effective sample size, and the unweighting efficiency,
thereby comparing the performance of Continuous Flows with that
of an unoptimised sample, with \Vegas,
and with Coupling Layer-based Normalizing Flows~\cite{Dinh:2014,Mueller:2019,Durkan:2019}.
We find that our new optimisation method
increases the effective sample size by factors between 8.3 and 70
for the highest multiplicities studied,
when compared to the conventional \Vegas-based optimisation.
Crucially, our method also yields large improvements in 
unweighting efficiency for the highest currently relevant multiplicity. 
To the best of our knowledge, we are the first to condition the phase-space sampler on the helicity states of the external particles, which allows us to optimise for individual helicity amplitudes and exploit correlations between helicities and kinematic variables. 
We provide a minimal interface to the \Pepper\ event
generator allowing for straight-forward integration of similar techniques in the future.

Our article is structured as follows: In section~\ref{sec:ps_in_hep}, we introduce the main problem we are trying to solve and bring it into the larger context of phase-space sampling in high-energy physics. We then provide a summary of Continuous Normalizing Flows
and Flow Matching in section~\ref{sec:cnfbg}. In section~\ref{sec:application} we describe our approach
for applying Flow Matching to phase-space sampling for relevant LHC processes using the \Pepper\ event generator. The results are then presented in section~\ref{sec:results},
followed by our conclusions in section~\ref{sec:conclusions}.
\section{Problem statement: Phase-space sampling}
\label{sec:ps_in_hep}
In this section, we give an overview of phase-space sampling in high-energy physics.
In order to make it more accessible to readers that are not domain experts,
we begin our description in section~\ref{sec:problem_statement} in mathematical terms by formulating the problem as an integral to be estimated by importance sampling. 
Section~\ref{sec:unweighting} introduces unweighted event generation as a different perspective motivated by typical HEP applications. In both cases, integration and unweighting, a trainable remapping from a uniform distribution to a nonuniform one can improve the relevant metrics. This is described in section~\ref{sec:remapping}, where we also derive the respective training objectives to be minimised. Finally, the connection of the mathematical treatment to physics and the discussion of domain-specific aspects
follow in section~\ref{sec:physics}.
\subsection{Phase-space sampling in mathematical terms}
\label{sec:problem_statement}
The main problem is to evaluate the integral
\begin{equation}
    I = \int_M \dif x \, p(x) f(x) 
\end{equation}
over a manifold \(M\). In our application, the variable \(x\) is a vector that describes a scattering event by incorporating the momentum components of the involved particles. The probability density function \(p\) describes the probability of a scattering event through the relation \(p(x) \dif x = \sigma^{-1} \dif \sigma\), where \(\sigma\) is the scattering cross section, and the function \(f\) gives the value of the physical observable.

Our approach to solve this problem is Monte Carlo integration using importance sampling. 
To make the problem more suitable for Monte Carlo sampling, we pull back \(p\) to the unit hypercube \(U = \intervalrightopen{0, 1}^d\) using a bijective map \(\phi: U \to M\), such that
\begin{equation}
    \label{eq:mc-map}
        I = \int_0^1 \dif^d x \, \det\biggl[{\pd{\phi}{x}}(x)\biggr] p\bigl(\phi(x)\bigr) f\bigl(\phi(x)\bigr)
          = \int_0^1 \dif^d x \, p_0(x) f\bigl(\phi(x)\bigr) \,,
\end{equation}
where we introduced the density \(p_0\) as the pull-back of \(p\) by \(\phi\):
\begin{equation}
    p_0(x) = \phi^* p(x) = p\bigl(\phi(x)\bigr) \det\biggl[{\pd{\phi}{x}}(x)\biggr] \,.
\end{equation}
In consequence, we can draw \(N\) points \(x_i\) independently from the multivariate uniform distribution \(u_d\) over \(U\) and approximate the integral \(I\) with the Monte Carlo estimator
\begin{equation}
    E = \frac{1}{N} \sum_{i=1}^N p_0(x_i) f\bigl(\phi(x_i)\bigr) = \frac{1}{N} \sum_{i=1}^N w_i \,.
\end{equation}
We call \(w_i\) the weight of the event with index \(i\). 
The expectation of \(E\) is \(I\) and its variance is given by
\begin{equation}\label{eq:importance_variance}
  \var(E) = \mathbb{E}(E^2) - \mathbb{E}(E)^2 \,.
\end{equation}
In a Monte Carlo simulation, it can be estimated by the unbiased sample variance. The variance is an indicator of the
sampling efficiency. When the variance is large, many points are needed to determine the integral with a given
precision. The size of the variance depends on the quality of the map \(\phi\).
\subsection{Unweighted event generation through rejection sampling}
\label{sec:unweighting}
A related task is the production of independent and identically distributed events following the distribution with density function \(p(x)\). In the context of HEP simulations, this is commonly called unweighted event generation. The need for it arises in the typical simulation toolchain used in experimental analyses. The parton-level event generation as described in section~\ref{sec:physics} is followed by further post-processing steps. These can be very expensive, especially the simulation of the detector response. If we were to use the points \(\phi(x_i)\) generated by importance sampling as described above, some of them would come with small weights. Running the post-processing on these events would consume significant resources, although their contribution to the integral is small. 
Consequently, generating unweighted parton-level events first is often more efficient, even though they are much more expensive to produce. Another advantage is that the sample is typically much smaller after unweighting, which saves storage space.

A general method to generate unweighted events is \emph{rejection sampling}~\cite{1961_Neumann_John}. It can be used in
the situation where it is desired to sample from the distribution \(p\) on \(M\) but one can originally only sample from
another distribution \(q\). In our case, we can sample from the uniform distribution \(u_d\) on \(U\) and apply the map
\(\phi\) to find \(q(x) = \phi_* u_d(x)\), 
where we introduced the push-forward operator
\begin{equation}\label{eq:push_forward_phi}
  \phi_* u_d(x) = u_d(x) \, \det\Bigl[ \pd{\phi^{-1}}{x} (x) \Bigr] = \det\Bigl[ \pd{\phi^{-1}}{x} (x) \Bigr] \,.
\end{equation}
Trial events are drawn from \(u_d\) and accepted or rejected based on an event-wise acceptance probability, given by
\begin{equation}\label{eq:is_weight}
    p_{\text{accept}}(x) = \frac{w(x)}{C} \quad \text{with  } w(x) = \frac{p_0(x)}{u_d(x)} = \frac{p(\phi(x))}{q(\phi(x))}\,,
\end{equation}
where \(C\) is a scaling constant chosen such that
\begin{equation}
    C \geq w(x) \quad \forall x \in M \,.
\end{equation}

This procedure generates a sample of reduced size with uniform weights \(w\),
with a distribution which is identical to the target distribution.
The resulting \emph{unweighting efficiency} 
is given by the average acceptance probability,
\begin{equation}\label{eq:unw_eff}
    \epsilon = \frac{\langle w\rangle}{C}\,,
\end{equation}
which is therefore a crucial figure of merit to measure the performance of a phase-space generator. The highest efficiency can be reached by choosing \(C=w_{\text{max}} \coloneq \max_x w(x)\). 
In practice, the maximal weight $w_{\mathrm{max}}$ needs to be estimated in an initial survey phase from a finite set of events.
Since the unweighting efficiency is very sensitive to single large outlier weights in this survey,
one often defines an effective $w_{\mathrm{max,eff}}$ to increase the efficiency and to render its value more numerically stable. 
When encountering events with weights $w>w_{\mathrm{max,eff}}$,
the event weight is set to $w/w_{\mathrm{max,eff}}$.
Such weights are referred to as overweights.
Choosing a suitable effective maximal weight requires balancing the introduced variance of overweights with the improved unweighting efficiency of a reduced maximal weight. 
A common approach is to define the desired fraction of overweight events via their contribution to the integral.
For example $\epsilon_{0.01}$ denotes the unweighting efficiency for which maximally 1\,\% of the integral
is contributed by events with weights that are larger than $w_{\mathrm{max,eff}}$.

Another frequently quoted measure to quantify the quality of a sample is the Kish effective sample size $N_{\mathrm{eff}}$, defined as
\begin{equation}
    \label{eq:neff}
    N_{\mathrm{eff}} = \frac{\langle w\rangle^2}{\langle w^2\rangle}\,.
\end{equation}
This estimates the relative size of the corresponding unit weight sample that would
result in the same variance of the integral estimator~\cite{kish1965survey}.
Note that $N_{\text{eff}}=1$ if all weights are equal,
i.e.\ if the weight distribution is given by $\delta(w - \langle w \rangle)$.
For a narrow weight distribution, the value of $N_{\text{eff}}$ will still be close to one,
while the value becomes smaller and approaches zero for wider distributions. In contrast to the unweighting efficiency, \(N_{\text{eff}}\) is less sensitive to large outlier weights.
\subsection{Remapping on the unit hypercube}
\label{sec:remapping}
\begin{figure}
    \centering
    \def\svgwidth{\columnwidth}
    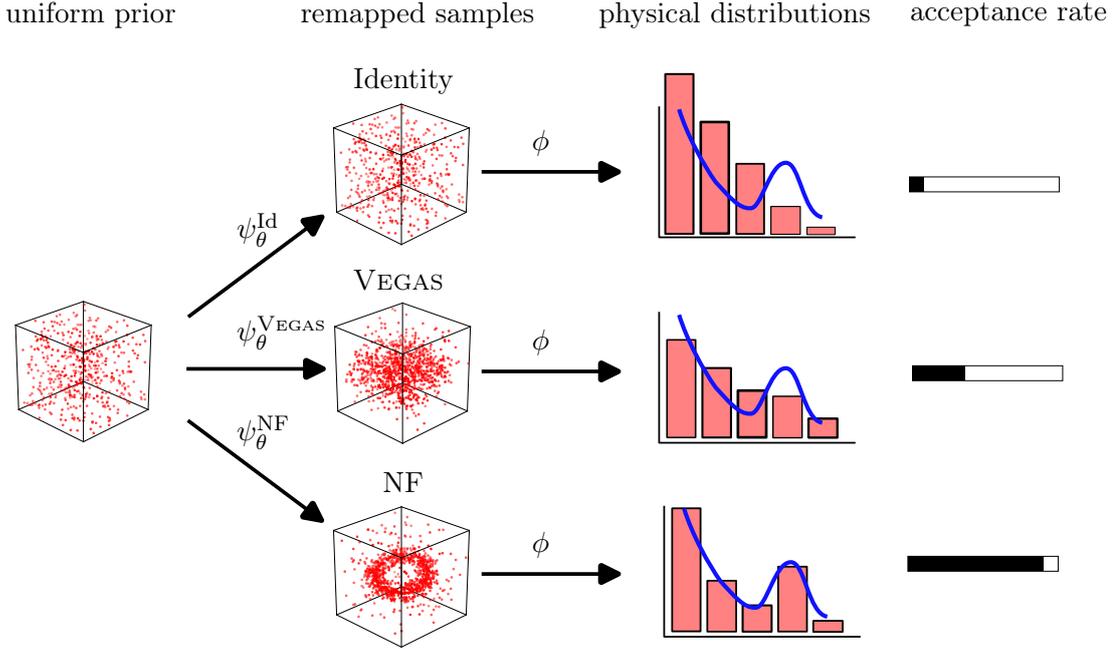
    \vspace{-1cm}
    \caption{Visualisation of the maps used in our integration approach.
    Beginning from the uniform prior, $\psi_\theta$ remaps the random numbers
and  $\phi$ maps to the physical variables. The latter are represented after a projection
onto a one-dimensional physical distribution, with the true distribution depicted as a solid line.
The Monte-Carlo acceptance rate depends on how similar the generated and the true distribution are.}
    \label{fig:mappings}
\end{figure}

All measures introduced above, the variance of the integral estimate, the unweighting efficiency, and the effective sample size, depend on the quality of the map \(\phi\). The map \(\phi\) can be constructed for specific scattering processes based on physics: A well-chosen \(\phi\) incorporates knowledge about the modes and general shape of \(p\) (and possibly \(f\)).
Although \(\phi\) can be designed manually for simple cases, finding an effective mapping becomes challenging for more complex scattering processes, especially those involving many particles. This leads to large variances in eq.~\ref{eq:importance_variance} and small unweighting efficiencies in eq.~\ref{eq:unw_eff}.
To address this problem, we use optimisation techniques and combine \(\phi\) with a second bijective map, \(\psi_{\theta}: U \to U\), a parametric model with parameters \(\theta\) that can be adapted to data.
This is visualised in figure~\ref{fig:mappings}.

We first consider the case of integration, as described in section~\ref{sec:problem_statement}. After introducing the map \(\psi_\theta\), the integral eq.~\ref{eq:mc-map} becomes
\begin{equation}
    \label{eq:optimised_integral}
    I = \int_0^1 \dif^d x \, \abs[\bigg]{\pd{\psi_{\theta}}{x}} p_0\bigl(\psi_{\theta}(x)\bigr) f\bigl(\phi\bigl(\psi_{\theta}(x)\bigr)\bigr)
\end{equation}
and its Monte Carlo estimator is now given by
\begin{equation}
    E_\theta = \frac{1}{N} \sum_{i=1}^N \abs[\bigg]{\pd{\psi_{\theta}}{x}} p_0\bigl(\psi_{\theta}(x_i)\bigr) f\bigl(\phi\bigl(\psi_{\theta}(x_i)\bigr)\bigr) \,.
\end{equation}
Its expectation value is \(\mean{E_\theta} = I\) by definition and its variance is
\begin{equation}\label{eq:var_E_theta}
  \var(E_\theta) = \mathbb{E}(E_\theta^2) - \mathbb{E}(E_\theta)^2 \,.
\end{equation}
Let
\begin{equation}\label{eq:push_forward_psi}
    q_\theta = (\psi_\theta)_* u_d \,.
\end{equation}
The variance, eq.~\ref{eq:var_E_theta}, becomes minimal when \(q_\theta\) is proportional to \(p_0 \cdot (f \circ \phi)\). Thus, if our goal is to minimise the variance, we need to optimise \(\psi_\theta\) such that \(q_\theta\) becomes close to \(p_0 \cdot (f \circ \phi)\). If the integral is the total scattering cross section \(\sigma\), the function \(f\) is constant and the task reduces to finding a \(q_\theta\) close to \(p_0\).

We can also consider unweighted event generation, cf.\ section~\ref{sec:unweighting}, with \(\psi_\theta\). In this case, we sample from \(q_\theta\) and accept events with probability
\begin{equation}\label{eq:is_weight_theta}
    p_{\text{accept}}(x) = \frac{w_\theta(x)}{C_\theta} \quad \text{with} \quad w_\theta(x) = \frac{p_0(x)}{q_\theta(x)} \,,
\end{equation}
where the ideal scaling factor is now \(C_\theta = \max_x w_\theta(x)\). The unweighting efficiency, eq.~\ref{eq:unw_eff}, becomes
\begin{equation}\label{eq:unw_eff_theta}
    \epsilon_\theta = \frac{\langle w_\theta\rangle}{C_\theta} \,.
\end{equation}
The highest unweighting efficiency is reached when \(q_\theta\) is equal to \(p_0\). In that case, all weights have a
value of one and all events are accepted. Therefore, in order to maximise the unweighting efficiency, we need to
optimise \(\psi_\theta\) such that \(q_\theta\) becomes close to \(p_0\). This is the same target as for reducing the
variance of the integral estimator in the case that \(f\) is constant. The precise metric is different, however. For
variance minimisation, we minimise the square loss \(\mathbb{E}(E_\theta^2)\) in eq.~\ref{eq:var_E_theta}, while for unweighting we minimise \(\max_x w_\theta(x)\), which determines the unweighting efficiency \(\epsilon_\theta\) in eq.~\ref{eq:unw_eff_theta}. In other words, for unweighting we are first and foremost interested in a reduction of the maximum weight, while the variance cares about the average deviation from the mean.
Note that in unweighting, it can be beneficial to intentionally overpopulate specific regions of phase-space.
This can happen, for example, when an event sample is meant to be used for several different analyses (looking at different observables \(f\)) and one of them needs a large number of events in the tail of some observable. If such an \enquote{enhancement} should be done with respect to an observable \(f\), it can be achieved by multiplying it to the event weight: \(w_\theta^{\text{enh}}(x) = f\bigl(\phi(x)\bigr) \cdot w_\theta(x)\). In that case, the optimal \(q_\theta\) is proportional to \(p_0 \cdot (f \circ \phi)\), just as for the integration case.

A basic example of 
\(\psi_\theta\)
is the Vegas algorithm, an adaptive importance sampler that factorises the problem into one-dimensional, piecewise-linear maps.
The key idea of \Vegas\ is to construct \(\psi_\theta\) as a product deformation map
\begin{equation}
    \psi_\theta^{\text{\Vegas}}(x) = \prod_{j=1}^d \psi_j(x_j) \,,
\end{equation}
where \(x_j\) denotes the \(j\)th component of the vector \(x\) and \(\prod\) is the Cartesian product. The one-dimensional maps \(\psi_j\) are piecewise linear and can be adapted by varying the widths of the linear segments. In a Monte-Carlo simulation, the adaption is typically performed in an initial optimisation phase, and the bin widths are optimised such that the variance is distributed equally among the bins.
The density \(q_\theta^{\text{\Vegas}} = (\psi_\theta^{\text{\Vegas}})_* u_d\) is piecewise constant. Therefore, sampling can be performed by selecting a bin randomly and sampling uniformly between the bin edges. 

While \Vegas\ can achieve large improvements in variance and unweighting efficiency in practice, it can fall short if \(p_0\) does not factorise. This can be alleviated by realising \(q(x) = \phi_* u_d(x)\) as a mixture density \(q(x) = \sum_i \alpha_i q_i(x)\), where the components \(q_i\) exploit known factorisations of individual terms contributing to the target function \(p(x)f(x)\). This \emph{multi-channel} approach, in combination with an optimisation of the mixture weights \(\alpha_i\)~\cite{Kleiss:1994qy} and a \Vegas\ remapping of each component \(q_i\)~\cite{Ohl:1998jn}, is the de facto state of the art in HEP. However, phase-space sampling is still a major challenge for complex scattering processes.

A more expressive alternative to \Vegas\ is given by Normalizing Flows~\cite{Tabak:2010,Tabak:2013,Dinh:2014},
which use neural networks to parameterise \(\psi_\theta\). 
Through restrictions in their architecture, Normalizing Flows can be designed in a way so that their Jacobian
determinant is tractable without having to invert the neural networks. Flows based on Coupling Layers~\cite{Dinh:2014}, for example,
implement the flow as a chain of discrete, simpler steps. In this work, we propose to realise \(\psi_{\theta}\) as a
Continuous Normalizing Flow~\cite{Chen:2018}, where the map is constructed implicitly by integrating a time-dependent vector field, which
is implemented by a neural network. We argue that Flow Matching~\cite{Lipman:2023,Albergo:2023building,Albergo:2023stochastic,Liu:2022} is a useful training method for the CNF, 
since it does not require to numerically solve an ODE during training. We compare this method with Normalizing Flows 
based on Coupling Layers, 
and with the \Vegas\ algorithm, 
which is the default method used in the \Pepper\ event generator and therefore serves as our baseline. 
We demonstrate in section~\ref{sec:results} that the CNF is stable to train and outperforms the other approaches for the
most challenging, i.e.\ highest dimensional, examples.
\subsection{Phase-space sampling in high-energy physics}
\label{sec:physics}
A central quantity in HEP simulations is the scattering cross section,
which is a direct measure of the probability of the given process to occur.
For hadronic collisions, e.g.\ at the LHC, it is given by
\begin{equation}
  \label{eq:hadronic_cross-section1}
  \sigma_{h_1h_2\to X} = \sum_{i,j} \int_0^1 \!\dif x_1 \int_0^1 \!\dif x_2 \  f_i(x_1,
  \mu_F)\,  f_j(x_2, \mu_F) \, \hat\sigma_{ij \to X}(x_1, x_2, \mu_R,\mu_F)\,,
\end{equation}
where the sum runs over all possible partons found within the incoming hadrons $h_{1,2}$, i.e.\ quarks, antiquarks, and
gluons. The individual partonic contributions (called subprocesses) can be integrated separately. This means we can
ignore the sum for now and concentrate on a single subprocess. 
The integration is performed over $x_{1,2}$, the light-cone momentum fractions of the partons $i,j$.
Finally, we have the renormalisation- and factorisation scales $\mu_{R,F}$ and the functions \(f_{i,j}\),
which are called parton distribution functions (PDFs). They can not be calculated from first principles and thus have to be determined from data. Typically, they are provided as precomputed grids in \(x\) and \(\mu_F^2\) and values between the knots are obtained using cubic interpolation. The PDFs are nonlinear and vary over several orders of magnitude.

The remaining ingredient of eq.~\eqref{eq:hadronic_cross-section1} is the partonic cross section \(\hat\sigma_{ij\to X}\), which is an integral over the final-state four-momenta $p_f = (E_f, \vec{p}_f)$, with $f=3,\ldots,m$ and $m$ being the number of incoming and outgoing particles. As a differential, it is given by
\begin{multline}
  \label{eq:qft_cross-section1}
  \mathrm{d}\hat\sigma_{ij\to X} = \frac{1}{2 E_1 E_2 \abs{\vec{v}_1-\vec{v}_2}} \, \Biggl( \prod_f
  \frac{\mathrm{d}^3 \vec{p}_f}{(2\pi)^3} \frac{1}{2E_f} \Biggr) \\
  \times \abs[\Big]{\symscr{M}_{ij\to X}\bigl(p_1,p_2 \to \{p_f\}\bigr)}^2 \, (2\pi)^4
  \, \delta^{(4)}\Biggl(p_1+p_2-\sum_{k=3}^{m} p_k\Biggr) \Theta\left(\left\{p_f\right\}\right) \,.
\end{multline}
We include the discontinuous cut function $\Theta$, which depends on the final-state momenta, and assumes either the value 0 or 1.
This implements the phase-space cuts, which are discussed in more details below.
The squared matrix element, \(\abs{\symscr{M}_{ij\to X}}^2\), is evaluated in perturbation theory, and is a complicated function that depends on the partonic process, $ij\to X$, and the incoming and outgoing particle momenta:
\begin{equation}
\label{eq:matrix-element1}
\abs[\Big]{\symscr{M}_{ij\to X}\bigl(p_1,p_2 \to \{p_f\}\bigr)}^2 \propto \sum_{\lambda_1,\dots,\lambda_d}
\symscr{A}^{\lambda_1,\dots,\lambda_m}(p_1,\dots,p_m) \symscr{A}^{\lambda_1,\dots,\lambda_m}(p_1,\dots,p_m)^\dagger\,,
\end{equation}
where most of the structure is expressed in the helicity amplitudes $\symscr{A}$. 
Here, the helicity amplitudes include the external polarisation vectors
and spinors for the external helicities $\lambda_i$,
and all required colour factors.
Certain helicity configurations $\lambda_1,\ldots,\lambda_d$
can result in vanishing $\mathcal A$.
These vanishing configurations can easily be identified
before the sampling begins and can then be ignored therein.
The number of non-vanishing helicity configuration
is referred to as $n_\text{hels}$ throughout this work.

The remaining component is the Lorentz-invariant phase space, which is needed to implement the symmetries of the system. These reduce the dimensionality of the partonic cross section from \(4n_\text{out}\) (with \(n_\text{out}=m-2\) being the number of outgoing particles) to \(3n_\text{out}-4\).
Together with the integrals over $x_{1,2}$, the total dimensionality is $d=3n_\text{out}-2$.
For the LHC, Monte-Carlo simulations typically have $n_\text{out} \lesssim 7$, such that $d \lesssim 19$.

The result of the integral for \(\sigma\) in eq.~\eqref{eq:hadronic_cross-section1}
is needed for overall normalisation, but can usually be estimated with sufficient accuracy without too much computational effort, at least at leading and next-to-leading order in perturbation theory.
The real computational challenge lies in generating large event samples distributed according to
\(\mathrm{d}\sigma\) in this high-dimensional phase space.
They have to contain sufficiently many events
to allow for flexible statistical comparisons
of distributions for a variety of physics observables,
both in the bulk and in the tails.
A good phase-space sampler must therefore be efficient throughout phase-space,
across regions where the integrand varies by many orders of magnitudes.

We can now connect this discussion with the generic description in section~\ref{sec:problem_statement}.
The manifold $M$ is given by all possible values for the momentum fractions $x_{1,2}$ and the final-state momenta $\{\vec{p}_f\}$,
where the fractions assume values between 0 and 1,
and the momenta are constrained by energy-momentum conservation
as implemented by the delta distribution in eq.~\eqref{eq:qft_cross-section1}.
The probability density function $p$ is given by
$p(x) \mathrm{d}x = \sigma^{-1} \mathrm{d} \sigma = \sigma^{-1} \mathrm{d}x_1 \mathrm{d}x_2 f_i f_j \mathrm{d}\hat{\sigma}$,
with the delta distribution removed.
Finally, in eq.~\eqref{eq:hadronic_cross-section1},
we consider the special case $f(x)=\sigma$.
However, in practice,
when evaluating the integral using Monte-Carlo integration,
one can store the generated sample of phase-space points
and the corresponding values of $p(x)$ and re-evaluate
the integral for any $f(x)$ in a fast and straightforward way.
This flexibility is one of the main advantages of using Monte-Carlo sampling.

Now, to find an optimal map $\psi_\theta$
in the sense defined in section~\ref{sec:remapping},
the task is to mimic the integrand of eq.~\eqref{eq:hadronic_cross-section1} closely,
to improve the estimate of the integral or the unweighting efficiency.
However, its structure makes this approximation challenging. 
Experimental and theoretical considerations require phase-space cuts $\Theta$ in eq.~\eqref{eq:qft_cross-section1},
effectively setting the matrix element to zero for certain regions.
These introduce discontinuities, often in regions where the integrand is large, since such cuts are commonly applied to remove singularities.
Furthermore, with the matrix elements $\abs{\symscr{M}}^2$ being comprised of products of strongly peaking helicity amplitudes,
see eq.~\eqref{eq:matrix-element1}, the integrand becomes multimodal, with potentially very narrow peaks separated from each other. 
Finally, the integration dimensions are strongly correlated. 
This correlation is potentially increased by the fact that, as indicated by the delta functions in eq.~\eqref{eq:qft_cross-section1}, the integration is only performed over a submanifold of reduced dimensionality (as defined by the Lorentz-invariant phase space). 
While this manifold can be mapped out efficiently, the mapping can introduce further correlations in the process.
\section{Background: Continuous Normalizing Flows}
\label{sec:cnfbg}
In this section, we describe how Continuous Normalizing Flows (CNFs) provide a viable solution to the sampling problem
described in section~\ref{sec:problem_statement}. We begin by introducing the concept of CNFs in section~\ref{sec:cnf}. There we also show how to generate samples with the flow and how to evaluate its density. In section~\ref{sec:flow_matching}, we show how a CNF can be trained efficiently, without solving the flow differential equation, using the Flow Matching method. We conclude in section~\ref{sec:conditioning} with a description of how to condition the flow on inputs with discrete values, like the helicity states of external particles in a scattering event.
\subsection{Flows from time-dependent vector fields}\label{sec:cnf}
As described in section~\ref{sec:ps_in_hep}, we want to generate samples from a target density\footnote{For simplicity in the notation, we here ignore the map \(\phi\) and drop the subscript \(0\)
accordingly. We also ignore the function \(f\) but note that it could be added to all equations with ease.} \(p: \mathbb{R}^d \to
\mathbb{R}\) but are only
able to sample from a latent density \(q_0\). We would like to find a map \(\psi: \mathbb{R}^d \to \mathbb{R}^d\) and
use it to transform samples \(x \sim q_0\) such that \(\psi(x) \sim p\), or at least that they are close in
distribution. Normalizing Flows provide such a map in the form of a parametrised diffeomorphism that can be trained on
data. Methods differ in how exactly the map is constructed and which loss functional is used for training. An example
are discrete-time flows based on a finite sequence of simple maps, \(\psi = \psi_k \circ \cdots \circ \psi_1\). These
can be built, for example, with Coupling Layers~\cite{Dinh:2014} and are typically trained with a Kullback--Leibler (KL)
loss. For phase space sampling in HEP, this option has been explored in
refs.~\cite{Bothmann:2020ywa,Gao:2020zvv,Stienen:2021gns,Verheyen:2022tov,Heimel:2022wyj,Heimel:2023ngj,Deutschmann:2024lml}. Below, we show how \(\psi\) can also be realised as a Continuous Normalizing Flow, which is the continuous-time analogue. As shown in section~\ref{sec:results}, we observe that CNFs are easier to train and scale better to higher dimensions than discrete-time flows.

Let \(v_t : [0, 1] \times \mathbb{R}^d \to \mathbb{R}^d\) be a smooth time-varying vector field. This vector field describes the velocity of a point \(x_t \in \mathbb{R}^d\) at time \(t \in [0, 1]\). The motion of the point is determined by the ordinary differential equation (ODE)
\begin{equation}\label{eq:ode}
    \od{x_t}{t} = v_t(x_t) \,.
\end{equation}
We denote the solution to this ODE by \(\psi_t : \mathbb{R}^d \to \mathbb{R}^d\) and the initial conditions are given by \(\psi_0(x_0) = x_0\) (or \(\psi_0 = \text{Id}\)). The map \(\psi_t\) is the time-dependent flow of the vector field \(v_t\). This can be used to generate samples at arbitrary times \(t\): We sample \(x_0 \sim q_0\) and apply the map \(\psi_t\), for which we have to integrate the ODE~\eqref{eq:ode}:
\begin{equation}\label{eq:cnf_sampling}
    x_t = \psi_t(x_0) = \int_0^t v_{t'}(\psi_{t'}(x_0)) \dif t' \,.
\end{equation}
This allows us to determine the trajectory of any point. By the Picard--Lindel\"of theorem, if $v_t$ is Lipschitz continuous in space and continuous in time, the solution to the ODE \eqref{eq:ode} is unique and trajectories of different initial points never intersect, which ensures that \(\psi_t\) has an inverse.

Another quantity we are interested in is the probability density \(q_t(x_t)\) for a generated point \(x_t\) at time \(t\). To derive this, we note that due to local conservation of probability, the density has to obey the continuity equation
\begin{equation}
    \pd{}{t} q_t(x) + \nabla \cdot (q_t(x) v_t(x)) = 0 \,.
\end{equation}
Plugging in \(x=x_t = \psi_t(x_0)\), cf.\ eq.~\eqref{eq:cnf_sampling}, it follows that
\begin{equation}
    \od{}{t} q_t(\psi_t(x_0)) + q_t(\psi_t(x_0)) \nabla \cdot v_t(\psi_t(x_0)) = 0 \,,
\end{equation}
which simplifies to
\begin{equation}
    \od{}{t} \log q_t(\psi_t(x_0)) + \nabla \cdot v_t(\psi_t(x_0)) = 0 \,.
\end{equation}
So to evaluate \(q_t(x_t)\), we need to solve this ODE by integrating it numerically from \num{0} to \(t\). In practice, sampling and density evaluation can be done jointly by solving the system
\begin{equation}\label{eq:joint_ode}
    \od{}{t} \begin{bmatrix}
           \psi_t(x_0) \\
           \log q_t(\psi_t(x_0))
         \end{bmatrix} = \begin{bmatrix}
                           v_t(\psi_t(x_0)) \\
                           - \nabla \cdot v_t(\psi_t(x_0))
                          \end{bmatrix}
\end{equation}
with initial conditions
\begin{equation}
    \begin{bmatrix}
        \psi_t(x_0) \\
        \log q_t(x_0)
    \end{bmatrix}_{t=0} = \begin{bmatrix}
                        x_0 \\
                        \log q_0(x_0)
                    \end{bmatrix} \,.
\end{equation}

We can also compute the density for an arbitrary point \(x_1\) at time \(t=1\), which has not necessarily been generated with the flow. Using an auxiliary function \(a_t\), the appropriate reverse ODE is
\begin{equation}\label{eq:joint_ode_reverse}
    \od{}{t} \begin{bmatrix}
           \tilde{\psi}_t(x_1) \\
           a_t
         \end{bmatrix} = \begin{bmatrix}
                           - v_t(\tilde{\psi}_t(x_1)) \\
                           \nabla \cdot v_t(\tilde{\psi}_t(x_1))
                          \end{bmatrix}
\end{equation}
and has be integrated from \(1\) to \(0\) with the initial conditions at t=1
\begin{equation}
    \begin{bmatrix}
        \tilde{\psi}_t(x_1) \\
        a_t
    \end{bmatrix}_{t=1} = \begin{bmatrix}
                        x_1 \\
                        0
                    \end{bmatrix} \,.
\end{equation}
The auxiliary function can then be used to determine the density:
\begin{equation}\label{eq:density_reverse}
    \log q_1(x_1) = \log q_0(x_0) - a_0 \,.
\end{equation}

Above, we have defined the domain of \(\psi_t\) to be \(\mathbb{R}^d\) instead of the unit hypercube \(U\) used in section~\ref{sec:remapping}. This is the natural way to define the flow, because it allows the vector field \(v_t\) to be unconstrained with regard to boundaries. In consequence, the natural (uninformed) base distribution is a standard normal. To map the output \(x_t\) of the flow to \(U\), we use the element-wise sigmoid transform:
\begin{equation}
    \operatorname{sig} : \mathbb{R} \to [0,1],\ x \mapsto \operatorname{sig}(x) = \frac{1}{1 + e^{-x}} \,.
\end{equation}
To account for the associated change in density, we have to multiply with the corresponding Jacobian determinant.

\subsection{Simulation-free training with Flow Matching}\label{sec:flow_matching}
So far we have discussed how the map \(\psi_t\) can be constructed from a velocity field \(v_t\) to generate samples \(x_t\) and how we can evaluate the density for a given point \(x_1\). What is still missing is a way to find a map \(\psi_t\) (or velocity field \(v_t\)) such that the density \(q_1(x)\) closely approximates the target density \(p(x)\). To this end, we use a model \(v_{t,\theta}\) with trainable parameters \(\theta\) to parametrise the vector field \(v_t\). A neural network is a convenient choice for \(v_{t,\theta}\), since eqs.~\eqref{eq:joint_ode} and~\eqref{eq:joint_ode_reverse} imply that for sampling or density evaluation, we only need to evaluate \(v_{t,\theta}\) and not its gradient or inverse. The vector field should be continuously differentiable, though.

Assume we are given training data in the form of samples \(x_1 \sim p(x)\). We want to use these data to adapt
\(v_{t,\theta}\). Given that we can evaluate the model density \(q_{t, \theta}\), using
eqs.~\eqref{eq:joint_ode_reverse} and~\eqref{eq:density_reverse}, an obvious way would be maximum likelihood estimation
(MLE), i.e.\ minimising the KL loss between \(q_{t, \theta}(x_1)\) and \(p(x_1)\), as proposed in~\cite{Chen:2018}. However, this approach has two major drawbacks: First, training is slow because in each step the ODE~\eqref{eq:joint_ode_reverse} has to be integrated with a numerical ODE solver, which needs many time steps to avoid large errors from the time discretisation. Second, there is no direct gradient information for the vector field for intermediate times between \num{0} and \num{1}, since only the endpoint of the flow is relevant for the KL loss. While there are approaches to regularise the flow, see e.g.\ refs.~\cite{Finlay:2020,Onken:2021}, these still require simulation during training.

The drawbacks of maximum likelihood training motivated the development of simulation-free training methods~\cite{Lipman:2023,Liu:2022,Albergo:2023building,Albergo:2023stochastic}. The idea is to directly match the vector field \(v_{t,\theta}\) to a target vector field \(u_t\), which generates the target density \(p\). There are various ways to construct an admissible \(u_t\). We briefly outline the strategy of refs.~\cite{Lipman:2023,Tong:2024} via conditional vector fields.
Let 
\begin{equation}
    u_t(x \mid x_0, x_1) = x_1 - x_0 \,,
\end{equation}
with \(x_0 \sim q_0\) and \(x_1 \sim p\) being samples from the base and target distribution, respectively. 
Clearly, a particle starting at \(x_0\) will flow to \(x_1\) along a straight line when following \(u_t(\cdot \mid x_0, x_1)\), i.e.\ 
\begin{equation}\label{eq:linear_interp}
    x_t = t x_1 + (1-t) x_0 \,.
\end{equation}
Let now
\begin{equation}\label{eq:marginal_target_vector_field}
    u_t(x) = \mathbb{E}_{\substack{(x_0, x_1) \sim \pi,\\ x_t=x}} [u_t(x \mid x_0, x_1)] \,.
\end{equation}
It can be shown~\cite{Lipman:2023,Tong:2024,Liu:2022} that the flow according to \(u_t\) transforms \(q_0\) into \(p\) as long as the marginals of the joint law \(\pi\) are \(q_0\) and \(p\). For instance, we can set \(\pi(x_0, x_1) = q_0(x_0) \cdot p(x_1)\) (independent coupling). Note that we condition the expectation to pairs \((x_0, x_1)\) where \(x_t\), eq.~\eqref{eq:linear_interp}, moves through the query point \(x\). 
It is now natural to fit \(v_{t, \theta}(x)\) to \(u_t(x)\), e.g.\ via
\begin{equation}\label{eq:FM_loss_marginal}
    \mathbb{E}_{\substack{(x_0,x_1) \sim \pi,\\ t \sim U_1}} \lVert v_{t,\theta}(x_t) - u_t(x_t) \rVert^2 \,.
\end{equation}
Note that evaluating \(u_t\) at specified points \(x\) is cumbersome due to the conditioning in eq.~\eqref{eq:marginal_target_vector_field}. Fortunately, one can show~\cite{Lipman:2023,Tong:2024} that eq.~\eqref{eq:FM_loss_marginal} has the same minimisers as
\begin{equation}\label{eq:FM_loss}
    \mathcal{L}_{\text{FM}} = \mathbb{E}_{\substack{(x_0, x_1) \sim \pi,\\ t \sim U_1}} \lVert v_{t,\theta}(x_t) -
    u_t(x_t \mid x_0, x_1) \rVert^2 \,.
\end{equation}
The flow matching objective, eq.~\eqref{eq:FM_loss}, has attractive properties. It is local in space and time, so it can be evaluated fast and without solving an ODE. Also, in contrast to a KL loss,
the minimiser is unique (on the support of the data). In consequence, sample-based training becomes easy.

More generally, one can add noise to the interpolation to increase robustness. This is done by adding noise with a small standard deviation \(\sigma_{\text{noise}}\) to the individual straight lines \(x_t\). This way, the sampled data cover more volume. For details, see refs.~\cite{Lipman:2023,Tong:2024}. 

We note that despite using a linear interpolation in eq.~\eqref{eq:linear_interp}, the flow trajectories \(x_t\) will in general not be straight lines. If the independent coupling is used, the samples \(x_0\) and \(x_1\) are paired randomly in eq.~\eqref{eq:FM_loss} and the vector field \(v_{t, \theta}\) will converge towards a local average of the interpolants. At time \(t=1\), it will still approximate the desired target \(p\) but when the trajectories are highly curved, sampling from the flow becomes slow because many small time steps are needed to achieve an accurate result with a numerical ODE solver. In the case of perfectly straight trajectories, a single step would suffice. Several approaches have been proposed to straighten the flow. In ref.~\cite{Liu:2022}, the authors propose an iterative procedure, where 
the first iteration uses the independent coupling and in each subsequent iteration, the coupling \(\pi\) is realised by the flow of the vector field \(v_{t, \theta}\) trained in the previous iteration. This means that \(x_0 \sim q_0\) and \(x_1 \sim q_1 = (\psi_\theta)_* q_0\) (which implies the method is not simulation-free). The authors show that each training renders the flow trajectories more straight. However, it also accumulates estimation error, since the samples are not drawn from \(p\) anymore. Therefore, the authors recommend to not apply too many iterations. Another approach is to use a coupling between \(x_0\) and \(x_1\) that already encourages straight trajectories during a single training iteration. This can be done by solving the optimal transport (OT) problem and drawing \(x_0\) and \(x_1\) from the OT plan \(\pi(x_0, x_1)\). Since this is computationally intractable for large datasets, one can instead use a minibatch approximation where the OT plan is computed only for minibatches of data. The resulting trajectories are not perfectly straight due to the bias from the approximation but it has been shown that even small batch sizes can lead to significant straightening~\cite{Tong:2024,Pooladian:2023}. In this work, we focus on the quality of the trained flows and do not investigate the use of straightening techniques. We note, however, that their application would be useful to reduce the sampling costs.

We make two modifications to the flow matching objective. First, in section~\ref{sec:problem_statement} we have used the map \(\phi\) to map our target distribution to the unit hypercube \(U\), which means that our training data is also defined on that domain. As noted in the previous section, the flow is defined on \(\mathbb{R}^d\). Our solution for sampling was to use a sigmoid transform. Here, we need to apply its inverse, the logit transform
\begin{equation}
    \operatorname{logit} : [0,1] \to \mathbb{R},\ x \mapsto \operatorname{logit}(x) = \ln \frac{x}{1 - x} \,,
\end{equation}
to map the training data to \(\mathbb{R}^d\). To avoid numerical issues when the input to the logit is large in magnitude, we compose it with an affine transformation, \(x \mapsto x \cdot (1-\epsilon) + \epsilon/2\). We found \(\epsilon = \num{1e-6}\) to work well in our case. For sampling, the inverse transformation has to be applied accordingly.

The second modification is related to the fact that, in contrast to what is assumed in eq.~\eqref{eq:FM_loss}, we do not sample \(x_1 \sim p\) since this is our goal in the first place. Instead, we sample from the uniform distribution \(u_d\). We can still use the flow matching objective by multiplying it with the importance weight \(w\), cf.\ eq.~\eqref{eq:is_weight}. This accounts for the mismatch in density. The approach is a variant of the Energy Conditional Flow Matching objective proposed in ref.~\cite{Tong:2024}. Of course, the importance sampling can lead to a high variance of the Monte Carlo estimator of eq.~\eqref{eq:FM_loss} due to fluctuations in the weights. This is especially problematic in high dimensions. To keep this under control, it is important to use a map \(\phi\) that reduces the variance to such an extent that model training becomes feasible. This requires a deep understanding of the physics involved. After training, the flow is itself a good sampler that can be used to generate new training data with a narrower distribution of weights. This suggests an iterative training scheme, where the flow trained in one iteration is used to generate the training data by sampling from \(q_\theta = (\psi_\theta)_* u_d\) for the next iteration with weights according to \eqref{eq:is_weight_theta}. We note that \Vegas\ could also be used to generate the data for the first iteration. Similarly, a Markov-Chain Monte-Carlo sampler could be used~\cite{LaCagnina:2024wcc}.
\subsection{Helicity conditioning}\label{sec:conditioning}
Normalizing Flows can be used to model conditional probability density functions. This is particularly interesting in
combination with the \Pepper\ event generator, which allows to evaluate the sum over squared helicity amplitudes in
eq.~\ref{eq:matrix-element1} through a Monte Carlo sampling of the non-vanishing helicity configurations. In this case, the sampling of the particle momenta can be
conditioned on the discrete helicity configuration corresponding to the helicity amplitude to be evaluated. Some of the helicity
configurations can be highly correlated and through the use of neural networks, Normalizing Flows can be expected to learn these
correlations well. This is a clear advantage in comparison to \Vegas, which would need to adapt to each helicity configuration
individually, with poor convergence expected for configurations with low probability. By default, \Pepper\ uses a single \Vegas\
map for all helicity configurations, which averages over their respective distributions. 

We note that, in principle, Normalizing Flows
could be conditioned on any variable. In ref.~\cite{Janssen:2025zke} this has been used for the sectors occurring in NNLO
QCD calculations with the sector-improved residue subtraction method. Other examples include the different partonic
channels in scattering processes with hadronic initial states and jet final states or the different multiplicities
contributing to inclusive cross sections, which we will both explore in the future. Here we describe the general
method. 

Let \(k\) be the number of different discrete states. Each state is assigned an integer value, denoted by \(h \in 1
\ldots k\). To condition a Continuous Normalizing Flow on \(h\), we use a conditional vector field \(v_{t,
\theta}(\cdot \mid h)\). By integrating the ODE, this can be used to sample from a conditional model density
\(q_{1,\theta}(\cdot \mid h)\). Since the vector field is implemented by a neural network, we have to use a continuous
representation of \(h\) to feed the values into the network. To this end, we use an embedding layer, which is trained
jointly with the vector field. The representation from the embedding layer is then concatenated to the other input
variables and fed into the input layer of the network. The non-vanishing helicity configurations themselves are sampled randomly, with weights
determined from training data by their respective contribution to the integral.

\section{Application to \LHC\ event sample production}
\label{sec:application}
Here we present our approach for applying
our novel neural network optimisation method
to single partonic channels in two standard-candle processes at the \LHC,
lepton and top pair production in association with additional jets,
at a centre-of-mass energy of $\sqrt s = \SI{13}{\TeV}$.
For both processes, we focus on the single partonic channel
that contributes most of the overall cross section,
which is $d \bar d \to e^+ e^- + ng$
for lepton pair production
and $g g \to t \bar t + ng$
for top pair production.

\subsection{\Chili\ phase-space mapping}
In this work, we make use of the \Chili\ phase-space mapping~\cite{Bothmann:2023siu}
to provide the map $\phi : U \to M$, see equation~\eqref{eq:mc-map}.
The phase space for the final state particle $f$ in eq.~\eqref{eq:qft_cross-section1}, can be expressed in terms of its transverse momentum $p_{f,\perp}$, rapidity $y_f$ and azimuthal angle $\varphi_f$ via
\begin{equation}
    \frac{\mathrm{d}^3 \vec{p}_f}{(2\pi)^3} \frac{1}{2E_f} = \frac{1}{16\pi^2}\dif p_{f,\perp}^2\dif y_f \frac{\dif \varphi_f}{2\pi}\,.
\end{equation}
Using this parametrisation for all final-state QCD partons allows common transverse momentum and rapidity cuts to be efficiently applied.
In the phase-space generation, we generate the transverse momenta according to $\dif p_{f,\perp}^2 / (2p_{\perp,c}+p_{f,\perp})^2$ where $p_{\perp,c}$ denotes an adjustable parameter used to effectively map out potential phase-space cuts.
The remaining variables are distributed uniformly, the rapidity between a maximum and minimum rapidity determined by external cuts and the azimuthal angle between 0 and $2\pi$. 

To satisfy energy-momentum conservation, we use the above mapping for all but the last final state QCD parton. 
The phase-space integration for the last final-state momentum is then performed in combination with the integration over the light-cone momentum fractions $\dif x_{i,j}$ in eq.~\eqref{eq:hadronic_cross-section1}. 
This combined integration allows to trivially satisfy energy-momentum conservation, by using the remaining final-state momentum to balance transverse momenta and the initial state momenta for the overall conservation. 
For vector-boson production, we add an additional $s$-channel propagator mapping according to the well known two-body decay formula~\cite{Bothmann:2023siu} with the virtuality of the intermediate particle distributed according to a Breit--Wigner distribution.

The key advantage of this mapping is its simplicity. 
The integrator consists of a single channel, while still performing similarly compared to more involved algorithms for a wide range of LHC processes~\cite{Bothmann:2023gew}. 
The single-channel setup is an ideal starting point for the combination with remapping techniques using a
trainable map
\(\psi_\theta\), as introduced in section~\ref{sec:remapping}, since no multi-channel and consequently more complicated
techniques such as~\cite{Kleiss:1994qy,Ohl:1998jn} need to be considered.
\subsection{\Pepper\ event generator}
We use
the Monte Carlo generator \Pepper~\cite{Bothmann:2021nch,Bothmann:2022itv,Bothmann:2023gew}
to perform the calculation of the perturbative scattering amplitudes.
\Pepper\ is a novel, portable parton-level event generator with GPU acceleration,
developed specifically for computationally expensive processes at the \LHC.

Momentum configurations are sampled
using \Pepper's internal implementation
of the \Chili\ phase-space generator~\cite{Bothmann:2023siu}.
While \Chili\ is comparatively simple and was thus easily ported to GPU~\cite{Bothmann:2021nch},
it has been shown to achieve a similar performance
as more complex established generators,
such as \Comix~\cite{Gleisberg:2008fv},
which is part of the \Sherpa\ framework~\cite{Gleisberg:2008ta,Sherpa:2019gpd,Sherpa:2024mfk}.

The matrix elements as given in formula \eqref{eq:matrix-element1}
are evaluated by summing over the colour quantum numbers,
using a minimal colour decomposition~\cite{Melia:2013bta,Melia:2013xok,Johansson:2015oia}.
The helicity quantum numbers are sampled instead.
To achieve a good sampling efficiency,
\Pepper\ normally samples them according to their relative contribution
to the overall variance of the Monte-Carlo sample with respect to the total cross section.
This optimisation is not normally correlated
with the optimisation of the phase-space sampling, and is essentially equivalent to 
an additional dimension in the \Vegas\ optimiser with the number of bins being equal 
to the number of non-vanishing helicities $n_\text{hels}$.
However, when applying the neural network based optimisation
we instead jointly optimise the phase-space and helicity sampling
and thus expose the optimiser
not only to the correlations
between the different phase-space variables,
but also between these variables and the helicity.

To write out parton-level events for training the external Flow models
(the \Vegas\ model is implemented and trained within \Pepper)
we utilise \Pepper's LHEH5 event output format.
The format is based on the HDF5 database library~\cite{hdf5},
accessed through the \HighFive\ header library \cite{highfive},
and has been specified in refs.~\cite{Hoeche:2019rti,Bothmann:2023ozs}.
We extend the specification for the purpose of this study
to include the random numbers for each event
which are required for the training,
i.e.\ for the phase-space point and the helicity configuration.

After model training (see section~\ref{sec:training}),
the generated random numbers for each event must be read in by \Pepper\ again
to evaluate the performance of the optimisation.
For this, we have added a corresponding reader to \Pepper,
which allows to use externally generated random numbers and phase-space weights
instead of internally generated random numbers and weights.

The new input/output features have been released with version 1.2 of \Pepper.
They allow for a very straightforward applications of external optimisation models
via simple file-based interfaces,
which are nevertheless flexible and scalable on HPC
due to HDF5's fast database structure and MPI support.
\subsection{Event generation parameters}
\label{sec:setup}
As discussed, the two partonic processes we study are
$d\bar{d}\rightarrow e^+e^-+ng$ (lepton pair production)
and $gg\rightarrow t\bar{t}+ng$ (top pair production)
at $\sqrt{s}=13\,$TeV.
In both cases,
we use the \NNPDF3.0 PDF set~\cite{NNPDF:2014otw} to evaluate the PDFs \(f_{i,j}\) in eq.~\eqref{eq:hadronic_cross-section1},
with the running strong coupling evaluated according to the set,
using the \LHAPDF6 library~\cite{Buckley:2014ana}.
The renormalisation and factorisation scales
are set to $\mu_R^2=\mu_F^2=H_T'^2/2$ for lepton pair production
and $\mu_R^2=\mu_F^2=H_T^2/2$ for top pair production~\cite{Bern:2013gka}.
The electroweak parameters are $\mathrm{sin}^2\theta_w= 0.23155$ and $\alpha = 1/128.80$.
We use the $Z$ and top masses $m_Z=91.1876\,$GeV, $m_t=m_{\bar{t}}=173.21\,$GeV
and set the $Z$~width as $\Gamma_Z=2.4952\,$GeV.
All remaining quarks are considered to be massless.

We apply the following conventional phase-space cuts.
For lepton-pair production,
we require the invariant mass to satisfy
$66\,\mathrm{GeV}\leq m_{e^+e^-} \leq 116\,\mathrm{GeV}$.
For both processes,
we require all massless final-state partons
to satisfy $p_{T,j}>20\,\mathrm{GeV}$, $|\eta_j| < 5$ and $\Delta R_{ij}> 0.4$,
for the transverse momentum $p_T$ and the pseudorapidity $\eta$ of parton $j$,
and for the pair distance $\Delta R$ between two partons $i$, $j$,
respectively.
\subsection{Model and training parameters}
\label{sec:training}
For all setups and multiplicities,
the \Vegas\ grids have 100 bins in each dimension. The bin widths are optimised
using 15 optimisation steps. The number of points is determined, such that the
total number of points for the highest multiplicity becomes roughly $3\cdot10^8$.
This results in $2.5\cdot10^5$ training points per non-vanishing helicity
configuration in the lepton pair production and $8\cdot10^4$ points
in the top-quark pair production case.

The NFs are optimised in 8 iterations, similar to how \Vegas\ is trained. In contrast to
refs.~\cite{Heimel:2022wyj,Akhound:2024}, we do not use a buffer but completely replace the data in each iteration. This
allows us to separate training and data production, so that both can be made efficient independently, and to use a
minimal interface to \Pepper, which only needs to read in random numbers and corresponding Jacobians and to write out
weights. For the first iteration, \(10^8\) weighted events are generated with \Pepper\
without using any remapping. In the subsequent iterations, the trained flows are used to generate \(10^7\) nonuniform random
numbers per iteration as inputs, followed by a calculation of the corresponding momenta and weights with \Pepper, which then serve as
training data. An embedding layer is used to provide a continuous
representation of the non-vanishing helicity configurations, with its width being equal to the number of such configurations $n_\text{hels}$. In all cases, this
representation is concatenated to the other input variables and fed into the input layer of each neural network.

For the Coupling Flows, 
the minimal number of Coupling Layers able to model all correlations between inputs, \(2 \lceil \log_2 d
\rceil\)~\cite{Gao:2020vdv}, is used. The implementation is based on the \textsc{Nflows} python
package~\cite{nflows:2020}, using rational-quadratic Coupling Layers~\cite{Durkan:2019} with 16 spline knots per
dimension. Between two Coupling Layers, the inputs are permuted with a binary permutation scheme. Each Coupling Transform is
parameterised by a multi layer perceptron (MLP) with ReLU activation functions. The widths of the hidden layers are
growing exponentially, such that each hidden layer has twice as many nodes as the preceding layer. Hidden layers are
added until the final one has more nodes than the output layer. The size of the MLPs is thus dynamic and can vary
between the Coupling Layers. For the Coupling Flows,
20\,\% of the training data are used for validation and early stopping based on the validation loss is used. This was
found necessary to avoid overfitting. The AdamW
optimiser~\cite{adamw:2019} is used with a batch size of \(2^{16}\) and its learning rate is decreased on plateaus. For
the lowest multiplicity processes, \(e^+e^-+1,2\) gluons and \(t\bar{t}+1,2\) gluons, the initial learning rate is set to \(10^{-3}\).
However, this leads to unstable training for the higher multiplicities, requiring to decrease the learning rate. For \(e^+e^-+3,4,5\) gluons and
\(t\bar{t}+3\) gluons, a value of \(10^{-4}\) is thus used. For \(t\bar{t}+4\) gluons it is further decreased to
\(10^{-5}\). 

For the ODE
Flows, each vector field is implemented as an MLP with 4 hidden
layers of width 512, using SELU activation functions. As proposed in ref.~\cite{Wildberger:2023}, the time \(t\) is
sampled from the power-law distribution \(p_\alpha(t) \propto t^{1/(1+\alpha)}\), with \(\alpha=1\). A 16-dimensional
Fourier feature mapping~\cite{Tancik:2020} is used to encode the time parameter and the encoding is concatenated to the other input
variables. The ODE Flows are trained for 600 epochs with a batch size of \(2^{19}\) using the AdamW
optimiser with a learning rate of \(10^{-3}\). Neither early stopping nor learning rate decay is used, as neither was not
found to improve the results. For ODE solving, necessary for sampling and density evaluation with the flow, the
DormandPrince45 solver~\cite{DORMAND198019} as implemented
in the \textsc{TorchDYN} python package~\cite{politorchdyn} is used. The absolute and relative tolerances of the
adaptive solver are set to \(10^{-4}\). 

For all three mappings, \Vegas, Coupling Flows and ODE Flows, the total numbers of trainable parameters for the different processes are given in
table~\ref{tab:model_parameters}. The numbers increase with the dimensionality and the number of helicity configurations.
Compared to the Normalizing Flows, \Vegas\ uses about three orders of magnitude less parameters. For the two kinds of
flows, the numbers are of the same order of magnitude. 

\SetTblrInner{rowsep=0.pt}
\begin{table}
    \centering
    \begin{tblr}{
        hline{1,Z} = {\heavyrulewidth},
        hline{3,8} = {\lightrulewidth},
        hline{2} = {5-7}{\cmidrulewidth},
        colspec = {@{}cS[table-format=1.0]S[table-format=2.0]S[table-format=3.0]S[table-format=4.0]S[table-format=7.0]S[table-format=7.0]@{}},
        row{1,2} = {guard},
        row{1} = {abovesep=2pt,belowsep=1pt},
        row{2} = {abovesep=1pt},
        row{3,8} = {abovesep=2pt},
        row{2,7,Z} = {belowsep=2pt},
        }
        Process                                             & $n$ & $d$ & $n_\text{hels}$ & \SetCell[c=3]{c} model parameters \\
                                                            &     &       &             & \Vegas & Coupling Flow & ODE Flow \\
        \SetCell[r=5]{} \(d\bar{d} \rightarrow e^+ e^-+ng\) &   1 &     7 &           8 &    700 &        616880 &  1078878 \\
                                                            &   2 &    10 &          16 &   1005 &       1438515 &  1086249 \\
                                                            &   3 &    13 &          32 &   1318 &       2450475 &  1098300 \\
                                                            &   4 &    16 &          64 &   1674 &       3563647 &  1120863 \\
                                                            &   5 &    19 &         128 &   2008 &       5215120 &  1169058 \\
        \SetCell[r=4]{} \(gg \rightarrow t\bar{t}+ng\)      &   1 &     7 &          32 &    724 &        614402 &  1092150 \\
                                                            &   2 &    10 &          64 &   1053 &       1571755 &  1114713 \\
                                                            &   3 &    13 &         128 &   1414 &       2411499 &  1162908 \\
                                                            &   4 &    16 &         256 &   1839 &       2844223 &  1280799 \\
    \end{tblr}
    \caption{\label{tab:model_parameters}
      Number of parameters for the different mappings. The parameters include both the number of parameters required for
  the phase-space remapping and the choice of the helicity configuration. We also give the number of final-state gluons $n$, the resulting phase-space dimensionality $d=3n_\text{out}-2$, and the number of non-vanishing helicity configurations $n_\text{hels}$.}
\end{table}
\subsection{Evaluation}
\label{sec:evaluation}
After finishing the training of the models,
we freeze their parameters and generate
samples of 10 million weighted events that pass the phase-space cuts
for each model and each final-state jet multiplicity
for both processes.
For each sample, we calculate the cut efficiency
(the fraction of events that pass the phase-space cuts),
the relative Monte-Carlo error of the cross-section integral,
the Kish effective sample size $N_{\mathrm{eff}}$, eq.~\eqref{eq:neff},
and the effective unweighting efficiencies $\epsilon_{0.01}$
and $\epsilon_{0.001}$.
This is repeated ten times with different random seeds.
Mean values and standard deviations for the different figures of merit
are then estimated from these repeated samples.
\section{Results}
\label{sec:results}
We present results for the mean and standard deviation of several quality metrics:
phase-space cut efficiency,
relative integration error, effective sample size $N_\text{eff}$
and unweighting efficiencies
$\epsilon_{0.01}$ and $\epsilon_{0.001}$.
The results are given in table~\ref{tab:results}
for both $d\bar d \to e^+e^-+ng$ and $gg \to t\bar t + ng$,
for the non-optimised mapping (``Identity''), a \Vegas-optimised mapping,
and mappings that are implemented as Coupling Flows and ODE Flows.
As discussed in section~\ref{sec:ps_in_hep},
these metrics capture different sampling efficiency aspects.
For the relative integration error, lower values are better,
while in all other cases, higher values are better.
We find that our newly proposed ODE Flow method has the best performance in almost all cases.
The only exception is the phase-space cut efficiency,
where all optimisation methods perform well,
with only small differences between them.
The results for the unweighting efficiencies differ far more significantly,
in particular for the computationally most difficult
highest gluon multiplicities $n$ in the study.
\SetTblrInner{rowsep=0.pt,colsep=5pt}
\begin{table}
    \sisetup{text-series-to-math = true, 
        detect-weight=true,
        mode=text
    }
    \centering
    \begin{tblr}{
        hline{1,Z} = {\heavyrulewidth},
        hline{2,22} = {\lightrulewidth},
        colspec = {@{}lS[table-format=1.0]lS[table-format=2.4(1)]S[table-format=1.6(1)]S[table-format=2.3(1)]S[table-format=2.4(1)]S[table-format=2.4(1)]@{}},
        row{1} = {guard,abovesep=2pt,belowsep=2pt},
        row{6,10,14,18,22,26,30,34} = {abovesep=4pt},
        row{2,22} = {abovesep=2pt},
        row{21,Z} = {belowsep=2pt},
        cell{2-Z}{1} = {c,font=\Large,cmd=\rotatebox{90}},
        cell{2-Z}{2} = {c},
        cell{4,8,11,15,19,24,28,32,36}{4} = {font=\bftab}, 
        cell{4,9,13,17,21,24,28,32,37}{5} = {font=\bftab}, 
        cell{4,9,13,17,21,24,28,32,37}{6} = {font=\bftab}, 
        cell{5,8,13,17,21,24,28,32,33,37}{7} = {font=\bftab}, 
        cell{5,9,13,17,21,24,29,33,37}{8} = {font=\bftab}, 
        }
        Process                                               & $n$               & mapping  & {cut eff./\si{\percent}} $\uparrow$ 
                                                                                                                  & {rel.\ err./\si{\percent}} $\downarrow$ 
                                                                                                                                          & {$N_{\mathrm{eff}}$/\si{\percent}} $\uparrow$ 
                                                                                                                                                              & {$\epsilon_{\mathrm{0.01}}$/\si{\percent}} $\uparrow$ 
                                                                                                                                                                                   & {$\epsilon_{\mathrm{0.001}}$/\si{\percent}} $\uparrow$ \\
        \SetCell[r=20]{} \(d\bar{d} \rightarrow e^+e^-+ng\)   & \SetCell[r=4]{} 1 & Identity & 86.88   \pm 0.02   & 0.0943   \pm 0.0001   & 11.651 \pm  0.008 &  3.840  \pm 0.006  &  3.00   \pm 0.02   \\
                                                              &                   & \Vegas   & 98.002  \pm 0.006  & 0.05453  \pm 0.00005  & 25.72  \pm  0.02  &  8.12   \pm 0.03   &  4.24   \pm 0.08   \\
                                                              &                   & Coupling & 99.927  \pm 0.001  & 0.0034   \pm 0.0002   & 99.1   \pm  0.2   & 78.68   \pm 0.05   & 49.4    \pm 0.3    \\
                                                              &                   & ODE      & 99.886  \pm 0.002  & 0.00364  \pm 0.00003  & 98.98  \pm  0.03  & 78.91   \pm 0.02   & 57.6    \pm 0.2    \\
                                                              & \SetCell[r=4]{} 2 & Identity & 75.55   \pm 0.02   & 0.247    \pm 0.006    &  2.2   \pm  0.2   &  0.23   \pm 0.02   &  0.08   \pm 0.03   \\
                                                              &                   & \Vegas   & 95.287  \pm 0.007  & 0.0949   \pm 0.0004   & 10.52  \pm  0.08  &  1.67   \pm 0.02   &  0.67   \pm 0.05   \\
                                                              &                   & Coupling & 99.450  \pm 0.002  & 0.02     \pm 0.04     & 79     \pm 28     & 48.7    \pm 0.6    & 13      \pm 5      \\
                                                              &                   & ODE      & 98.383  \pm 0.006  & 0.00990  \pm 0.00009  & 92.8   \pm  0.2   & 35.0    \pm 0.2    & 15.5    \pm 0.2    \\
                                                              & \SetCell[r=4]{} 3 & Identity & 65.04   \pm 0.02   & 1.4      \pm 0.9      &  0.15  \pm  0.08  &  0.007  \pm 0.007  &  0.003  \pm 0.002  \\
                                                              &                   & \Vegas   & 92.75   \pm 0.02   & 0.175    \pm 0.003    &  3.4   \pm  0.1   &  0.32   \pm 0.02   &  0.10   \pm 0.03   \\
                                                              &                   & Coupling & 92.48   \pm 0.02   & 0.030    \pm 0.007    & 58     \pm  9     & 15.5    \pm 0.2    &  3.0    \pm 0.7    \\
                                                              &                   & ODE      & 90.60   \pm 0.01   & 0.02085  \pm 0.00006  & 77.2   \pm  0.2   & 22.28   \pm 0.06   & 11.0    \pm 0.2    \\
                                                              & \SetCell[r=4]{} 4 & Identity & 55.18   \pm 0.02   & 2.8      \pm 0.9      &  0.02  \pm  0.01  &  0.0011 \pm 0.0006 &  0.0011 \pm 0.0006 \\
                                                              &                   & \Vegas   & 88.87   \pm 0.02   & 0.32     \pm 0.02     &  1.1   \pm  0.2   &  0.07   \pm 0.01   &  0.02   \pm 0.01   \\
                                                              &                   & Coupling & 85.24   \pm 0.01   & 0.053    \pm 0.003    & 31     \pm  2     &  4.43   \pm 0.07   &  1.1    \pm 0.2    \\
                                                              &                   & ODE      & 83.099  \pm 0.009  & 0.0386   \pm 0.0002   & 48.7   \pm  0.3   &  8.96   \pm 0.04   &  3.95   \pm 0.09   \\
                                                              & \SetCell[r=4]{} 5 & Identity & 46.09   \pm 0.02   & 8        \pm 7        &  0.004 \pm  0.003 &  0.0004 \pm 0.0002 &  0.0004 \pm 0.0002 \\
                                                              &                   & \Vegas   & 84.29   \pm 0.01   & 0.52     \pm 0.04     &  0.4   \pm  0.1   &  0.024  \pm 0.008  &  0.007  \pm 0.003  \\
                                                              &                   & Coupling & 82.90   \pm 0.02   & 0.2      \pm 0.1      &  4     \pm  2     &  0.29   \pm 0.09   &  0.03   \pm 0.03   \\
                                                              &                   & ODE      & 77.69   \pm 0.02   & 0.061    \pm 0.004    & 28     \pm  2     &  3.68   \pm 0.05   &  1.29   \pm 0.08   \\
        \SetCell[r=16]{} \(gg \rightarrow t\bar{t}+ng\)       & \SetCell[r=4]{} 1 & Identity & 77.70   \pm 0.02   & 0.2352   \pm 0.0006   &  2.29  \pm 0.02   &  0.434  \pm 0.006  &  0.27   \pm 0.02   \\
                                                              &                   & \Vegas   & 98.368  \pm 0.004  & 0.05358  \pm 0.00006  & 26.27  \pm 0.04   &  7.10   \pm 0.05   &  3.2    \pm 0.1    \\
                                                              &                   & Coupling & 99.9943 \pm 0.0003 & 0.00466  \pm 0.00003  & 97.92  \pm  0.03   & 67.17   \pm 0.04   & 45.6    \pm 0.3    \\
                                                              &                   & ODE      & 99.9843 \pm 0.0004 & 0.00515  \pm 0.00005  & 97.47  \pm  0.05  & 65.21   \pm 0.03   & 45.50   \pm 0.09   \\
                                                              & \SetCell[r=4]{} 2 & Identity & 65.40   \pm 0.02   & 0.47     \pm 0.02     &  0.68  \pm  0.03  &  0.063  \pm 0.005  &  0.021  \pm 0.007  \\
                                                              &                   & \Vegas   & 95.997  \pm 0.006  & 0.0676   \pm 0.0002   & 18.70  \pm  0.04  &  3.55   \pm 0.02   &  1.49   \pm 0.04   \\
                                                              &                   & Coupling & 99.205  \pm 0.003  & 0.013    \pm 0.002    & 85     \pm  2     & 35.9    \pm 0.2    & 10.2    \pm 0.9    \\
                                                              &                   & ODE      & 98.328  \pm 0.005  & 0.0205   \pm 0.0003   & 71.7   \pm  0.5   & 24.39   \pm 0.04   & 13.35   \pm 0.08   \\
                                                              & \SetCell[r=4]{} 3 & Identity & 54.27   \pm 0.03   & 1.1      \pm 0.2      &  0.13  \pm  0.03  &  0.008  \pm 0.003  &  0.003  \pm 0.001  \\
                                                              &                   & \Vegas   & 92.640  \pm 0.009  & 0.0904   \pm 0.0003   & 11.79  \pm  0.08  &  1.60   \pm 0.02   &  0.59   \pm 0.04   \\
                                                              &                   & Coupling & 94.008  \pm 0.006  & 0.0246   \pm 0.0002   & 66.4   \pm  0.4   & 15.26   \pm 0.06   &  6.3    \pm 0.3    \\
                                                              &                   & ODE      & 93.68   \pm 0.02   & 0.02768  \pm 0.00004  & 60.49  \pm  0.06  & 15.26   \pm 0.03   &  8.7    \pm 0.2    \\
                                                              & \SetCell[r=4]{} 4 & Identity & 44.36   \pm 0.02   & 3        \pm 1        &  0.03  \pm  0.02  &  0.0011 \pm 0.0008 &  0.0010 \pm 0.0007 \\
                                                              &                   & \Vegas   & 88.304  \pm 0.008  & 0.123    \pm 0.001    &  7.0   \pm  0.1   &  0.74   \pm 0.02   &  0.23   \pm 0.03   \\
                                                              &                   & Coupling & 91.269  \pm 0.005  & 0.2      \pm 0.2      &  8     \pm  6     &  1.6    \pm 0.5    &  0.04   \pm 0.06   \\
                                                              &                   & ODE      & 88.456  \pm 0.008  & 0.0300   \pm 0.0002   & 59.6   \pm  0.5   & 12.53   \pm 0.03   &  5.76   \pm 0.09   \\
    \end{tblr}
    \caption{\label{tab:results}
    Comparison of different mappings for the $gg \to t\bar{t}+ng$ example using samples with 10M nonzero weight events.
  The relative error of the total cross section estimate has been determined from 10M events including zero weights. The
uncertainties have been obtained as the standard deviation of 10 independent samples generated with the same mapping
(without re-training). All errors and uncertainties are rounded up in the last digit given.}
\end{table}

For $d\bar d \to e^+e^-+ng$ with $n=5$
we find $\epsilon_{0.01} = \SI{3.68(5)}{\percent}$ for the ODE Flow,
which is about 13 times higher than for the Coupling Flow’s
$\SI{0.29(9)}{\percent}$
and about 150 times higher compared
to \Vegas’ $\SI{0.024(8)}{\percent}$.
For $\epsilon_{0.001}$,
we find even larger improvements
of 43 and 184 when comparing to the Coupling Flow
and to \Vegas, respectively.

For $gg \to t\bar t +ng$ with $n=4$,
the ODE Flow's efficiency is $\SI{12.53(3)}{\percent}$,
which is 8 times higher
than the Coupling Flow's \SI{1.6(5)}{\percent},
and about 17 times higher compared to \Vegas' efficiency,
$\SI{0.74(2)}{\percent}$.
For $\epsilon_{0.001}$,
the improvements over the Coupling Flow
and \Vegas\ are 144 and 23, respectively,
with the Coupling Flow's efficiency falling below \Vegas' efficiency
in this case.

The unweighting efficiency improvements of the Flow-based methods
over \Vegas\ are consistently
higher for $d\bar d \to e^+e^-+ng$ when compared to $gg \to t\bar t +ng$
at the same $n$. For $n=4$, we find a factor of 128 for the former
and a factor of 17 for the latter.
The main reason is that \Vegas\ performs significantly better
for $gg \to t\bar t +ng$, for all $n$.
This is not only true for the unweighting efficiency $\epsilon$,
but to a similar degree also for the relative Monte-Carlo error,
indicating that this can not be traced back to the optimisation target of \Vegas.
An explanation would be that the prerequisite of \Vegas---i.e.\ that the distribution is factorisable into each of its dimensions---is fulfilled to a lesser degree for $d\bar d \to e^+e^-+ng$.
Here, \Chili\ introduces an $s$-channel for the $Z$-boson resonance,
and there is a more pronounced dependence on the helicity configuration.

In terms of the relative integration error, we find similar improvements. The Normalizing Flows achieve significant
improvements for all multiplicities considered. For $d\bar d \to e^+e^-+ng$ with $n=5$ we find
a value of $\SI{.061(4)}{\percent}$ for the ODE Flow, which is about 3 times less than the Coupling Flow and about 9
times less than \Vegas. Considering that the integral error scales as $1/\sqrt{N}$ we need to square these numbers to
determine the improvement in efficiency, i.e.\ the reduction in the number of points needed to determine the integral
with the same error. This results in factors of 11 and 73, respectively, in comparison to the Coupling Flow and \Vegas.
The corresponding factors for $gg \to t\bar t +ng$ with $n=4$ are 44 and 17, where in analogy to the results for the
unweighting efficiency the Coupling Flow produces a larger error than \Vegas. Over all multiplicities, the improvements
range between 14 and 224.

The effective sample size also benefits significantly from the Normalizing Flows. While the identity mapping achieves
values of $N_{\text{eff}} > \SI{1}{\percent}$ only for the lowest multiplicities, \Vegas\ achieves this in all cases except $d\bar d \to
e^+e^-+5g$. The Normalizing Flows, however, lead to much higher values and with the ODE Flow the effective sample size
is above $\SI{25}{\percent}$ in all cases considered. In accordance with the results for the unweighting efficiency and the relative
integration error, the highest multiplicities favour the ODE Flow over the Coupling Flow. For the lowest multiplicity,
$n=1$, the effective sample size with the Normalizing Flows is close to $\SI{100}{\percent}$. Accordingly, the
improvement with respect to \Vegas, which achieves about $\SI{25}{\percent}$, is close to the maximum factor of about 4.
The improvement factor increases with multiplicity and is highest for $d\bar d \to e^+e^-+5g$, where the ODE Flow
produces a 70 times larger effective sample size in comparison to \Vegas. With respect to the Coupling Flow, the
factor is about 9 in this case.

\begin{figure}[tbp]
    \centering
    \includegraphics[width=.48\textwidth]{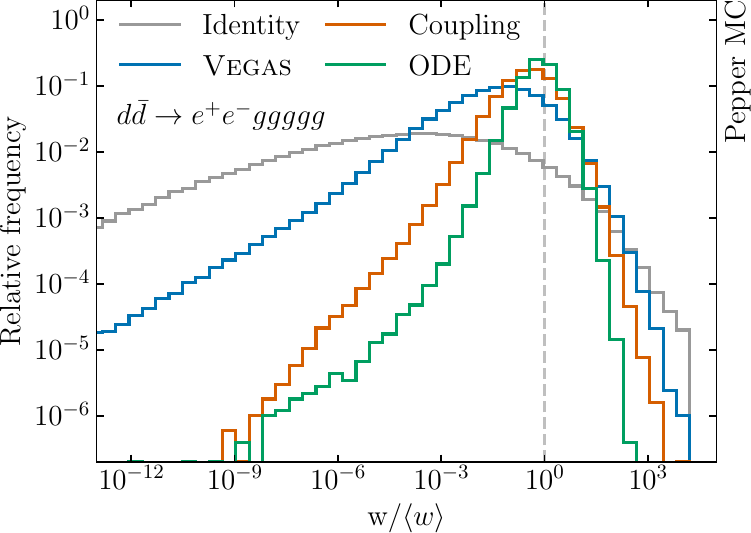}\hfill
    \includegraphics[width=.48\textwidth]{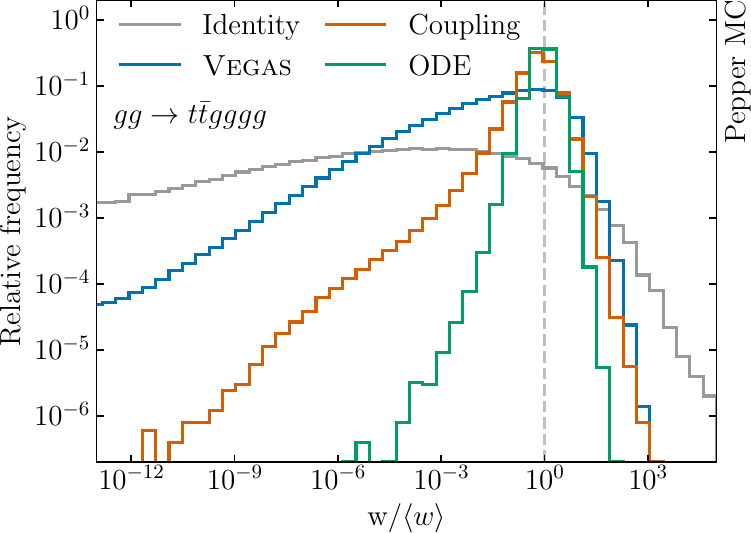}
    \caption{\label{fig:weight_dist}
    Normalised weight distributions for $5\cdot10^6$ points for the \(e^+e^-+5\) gluons example (left) and the $t\bar{t}$+4
    gluons example (right). The plotted weights include matrix-element and phase-space weights, as well as PDF and, where
    applicable, the Jacobian of the remapping.
    Distributions are shown for the ``Identity'' mapping,
    and mappings optimised by \Vegas,
    Coupling Flows (``Coupling''),
    and ODE Flows (``ODE'').}
\end{figure}

The improved sampling by the ODE Flow method also manifests itself
in a narrower weight distribution, as shown in figure~\ref{fig:weight_dist}
for $d\bar d \to e^+e^-+ng$ with $n=5$ (left)
and $gg \to t\bar t +ng$ with $n=4$ (right).
A narrower distribution is directly
related to a larger effective sample size $N_\text{eff}$.
As discussed in section~\ref{sec:remapping},
for a good unweighting efficiency 
the most relevant feature of the weight distribution is the steepness
of its tail towards large values $w$,
since this will determine the value of the denominator, $w_\text{max,eff}$, in eq.~\eqref{eq:unw_eff}.
Indeed, the ODE Flow yields a significantly steeper right tail in the weight distribution for both cases, compared to the other methods
including the Coupling Flows.

\begin{figure}[tbp]
    \centering
    \includegraphics[width=.48\textwidth]{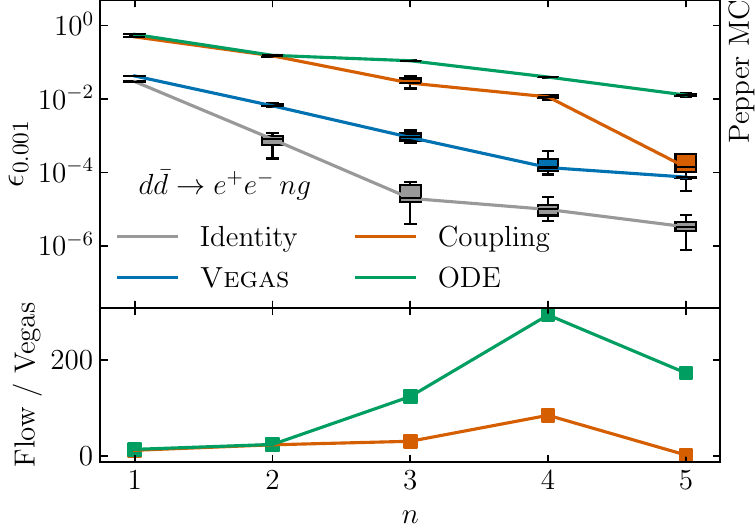}\hfill
    \includegraphics[width=.48\textwidth]{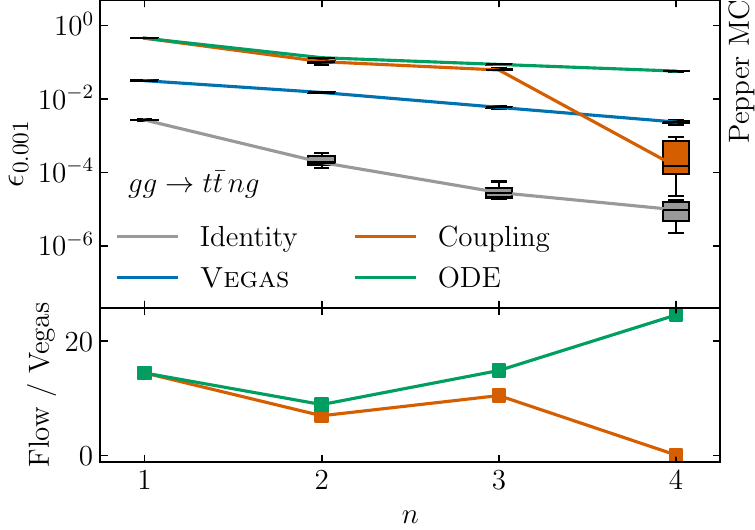}
    \caption{\label{fig:unweff}
        Unweighting efficiencies $\epsilon_{0.001}$ for the \(e^+e^-\) + $n$ gluons
        example (left) and the $t\bar{t}$+$n$ gluons example (right),
        for various gluon multiplicities $n$.
        The efficiencies are shown for the ``Identity'' mapping,
        and mappings optimised by \Vegas,
        Coupling Flows (``Coupling''),
        and ODE Flows (``ODE''). Each curve represents the mean over ten independent evaluations of the unweighting efficiency, the boxes indicate the quartiles of the distribution and the whiskers show $\sfrac{3}{2}$ of the interquartile range.
    }
\end{figure}

To study the scaling behaviour with the number of final-state gluons $n$,
we compare the values of $\epsilon_{0.001}$ in figure~\ref{fig:unweff}.
We find that the ratio of the efficiency for the ODE Flow over the efficiency of \Vegas\
generally increases with $n$.
On the other hand,
the Coupling Flows at first show similar improvements over \Vegas,
but for the largest $n$ decrease in efficiency significantly,
for both processes,
indicating an unfavourable scaling behaviour towards large~$n$.
By evaluating the training results ten times with different random seeds,
we can plot spreads for the data points.
We find that the they are smallest for the ODE Flows,
which means that this method generates
the fewest very large event weight outliers.

\begin{figure}[tbp]
    \centering
    \includegraphics[width=.48\textwidth]{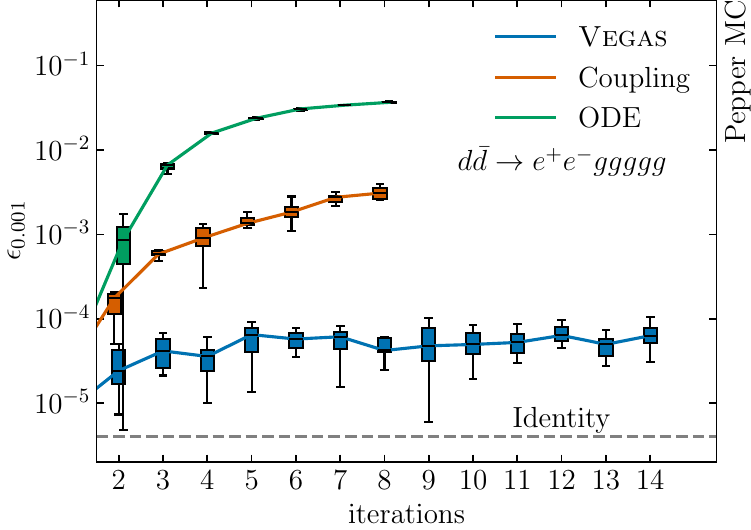}
    \hfill
    \includegraphics[width=.48\textwidth]{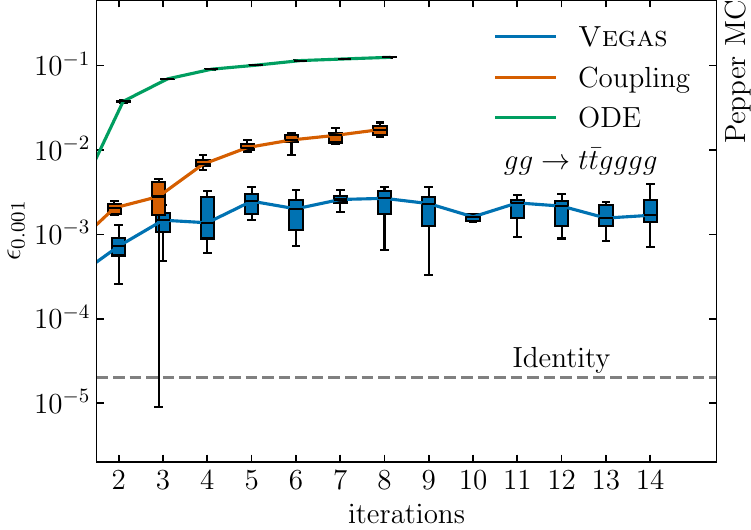}
    \caption{\label{fig:convergence}
        Convergence of the effective unweighting efficiency $\epsilon_{\mathrm{0.001}}$
        with training iterations for the highest final-state gluon multiplicities studied: 
        $d\bar d \to e^+e^-+5g$ (left) and 
        $gg \to t\bar t+4g$ (right). Results are shown for the ``Identity'' mapping, Vegas, Coupling Flow (``Coupling''), and ODE Flow (``ODE''). Each curve represents the mean over ten independent evaluations of the unweighting efficiency, the boxes indicate the quartiles of the distribution and the whiskers show $\sfrac{3}{2}$ of the interquartile range.
    }
\end{figure}

Finally, we analyse the convergence behaviour of $\epsilon_{0.001}$
  as a function of training iterations
in figure \ref{fig:convergence}
for the highest multiplicity of each process.
We find that all methods are reasonably converged at their last iteration.
The spread of their results over ten independent samples
is consistently the smallest for the ODE Flows towards later iterations,
with good stability from one optimisation step to the next.
We observe that the \Vegas\ results are less stable for
$d\bar d \to e^+e^-+ng$ when compared to $gg \to t\bar t +ng$,
consistent with the lower overall performance we have observed earlier
for the former process.
Also the Coupling Flows give numerically less stable results
when compared to the ODE Flows,
with a larger spread of results,
indicating a larger fraction of outliers with large weights affecting the
$w_\text{max,eff}$ determination. This might be an effect of the Coupling Flow's MLE objective versus the regression-based Flow Matching objective.

The above observations support the conclusion
that the new Flow Matching based method performs best,
in particular in the most relevant metrics
for event generation,
the unweighting efficiency.
The advantage over the Coupling Flow based approach
is most pronounced at the highest multiplicities studied,
where the performance of the Coupling Flows drops significantly. This observation is consistent with
others, which empirically show that time-discrete flows trained with MLE fall short in terms of
expressivity and scalability~\cite{Rehman:2025}. 
Since the highest multiplicities drive the computational cost
of event generation for current state-of-the-art production campaigns
at the LHC,
using a Flow Matching based approach would be highly beneficial.
\section{Conclusion}
\label{sec:conclusions}
We have presented the first application of the Flow Matching method
to the problem of high-dimensional phase-space sampling in high-energy particle physics.
The model is trained not only on the continuous kinematic phase-space dimensions,
but also on the discrete variable used to select a helicity configuration.
The training of the model and testing it in production is facilitated
by a newly implemented simple file-based interface in the event generator \Pepper.
The interface builds on the LHEH5 parton-level event database format,
and is used to train the model and then test it in production.
The interface is simple
and can be used in the future
to study more applications and also other novel sampling methods
in a straightforward way.

In this contribution, we studied Drell--Yan and top--antitop pair production.
Specifically, we evaluated various sampling efficiency metrics for the
most challenging partonic channels $d\bar d \to e^+e^-+ng$ with $1 \leq n \leq 5$
and $gg \to t\bar t+ng$ with $1 \leq n \leq 4$.
The upper limits of these multiplicity ranges correspond
to the highest number of jets currently simulated
at the matrix-element level for experimental analyses.
Their high complexity and low phase-space efficiency
render them the main bottlenecks
in current large-scale state-of-the-art event generation campaigns.
The key performance metric for these productions
is the unweighting efficiency $\epsilon_{0.01}$.
Here, we find gain factors of 150 (17)
for $d\bar d \to e^+e^-+ng$ with $n=5$
($gg \to t\bar t+ng$ with $n=4$).
These substantial improvements can be directly translated into increased
parton-level event generation throughput.
Overall, these results represent a significant advancement in the scalability
and efficiency of event generation for experimental physics analyses.

This study focused on the computationally most expensive partonic channel
of each process.
However, realistic simulations require the inclusion of all partonic channels during both training and generation.
In a future work, we aim to develop a single conditional model that can
simultaneously sample from all partonic channels.
This combined approach enables the model to leverage correlations between
channels, improving training effectiveness.
Similar correlations also exist across different jet multiplicities.
Consequently, a complementary strategy is to investigate the joint training
of all multiplicities within a single model to further enhance training
efficiency and production performance. Another improvement would be the use of rectification
methods~\cite{Liu:2022,Tong:2024,Pooladian:2023} that straighten the flow trajectories. These have the potential to
significantly reduce the cost of the Flow sampling.
\acknowledgments
The authors gratefully acknowledge the computing time granted by the Resource Allocation Board and provided on the supercomputer Emmy/Grete at NHR-Nord@Göttingen as part of the NHR infrastructure. The calculations for this research were conducted with computing resources under the project nhr\_ni\_starter\_22045.
The authors also acknowledge the use of computing resources made available
by CERN to conduct some of the research reported in this work.
This material is based upon work supported by Fermi Forward Discovery Group, LLC under Contract No.\ 89243024CSC000002 with the U.S.\ Department of Energy, Office of Science, Office of High Energy Physics.
EB and MK acknowledge support by the Deutsche Forschungsgemeinschaft
(DFG, German Research Foundation) -- 510810461. TJ acknowledges financial support from the German Federal Ministry of Education and
Research (BMBF) in the ErUM-Data action plan through the KISS consortium (Verbund-
projekt 05D2022). 
BS and FHS were supported by the German Research Foundation (DFG) SFB 1456, Mathematics of Experiment -- Project-ID 432680300. BS was supported by the Emmy Noether Programme of the DFG -- Project-ID 403056140.

\appendix
\section{Event generation runcards}
\label{app:run_cards}
The \Pepper\ v1.3 event generator\footnote{%
\url{https://gitlab.com/spice-mc/pepper/-/releases/1.3.0-kokkos}}
runcards that were used for this project
are given in listings~\ref{listing:runcard:dy} and~\ref{listing:runcard:tt}.
In addition to using the runcards,
the code needs to be compiled after changing the line
\texttt{\#define HELICITY\_BLOCK\_SIZE~32}
to
\texttt{\#define HELICITY\_BLOCK\_SIZE~1}
in \texttt{src/event\_data.h},
to avoid the generation of blocks of correlated events
that share a given helicity configuration.

The runcards were used to generate
the \Vegas\ and non-optimised ``Identity'' baseline results,
to generate the training data for the Flow models
and then to measure the performance of the trained models.
For the latter, the runcard line for the \texttt{input\_path} setting should be used
(by removing the semicolon to uncomment the line),
to enable reading the random numbers generated by the model,
instead of relying on \Pepper's internal random-number generator.

\begin{listing}
\begin{minted}[bgcolor=listingsbg,fontsize=\footnotesize]{ini}
[main]
process = d db -> e+ e- g g g g g
mu2 = H_Tp^2/2
; helicity_integrator = false

[events]
random_numbers_output_enabled = true
unweighting_disabled = true
output_path = training_data_for_flow_model.hdf5
; input_path = random_numbers_from_flow_model.hdf5

[phase_space.vegas]
; disabled = true
alpha = 0.8

[phase_space.optimisation]
n_iter = 15
n_nonzero_min = 250000
n_nonzero_min_growth_factor = 1.00
\end{minted}
\caption{\label{listing:runcard:dy}
	The \Pepper\ runcard used for generating $d\bar{d} \to e^+e^- ggggg$ events.
	For generating a different number of final-state gluons, only the \texttt{process} setting
	needs to be adjusted.}
\end{listing}
\begin{listing}
\begin{minted}[bgcolor=listingsbg,fontsize=\footnotesize]{ini}
[main]
process = g g -> t tb g g g g
mu2 = H_T^2/2
; helicity_integrator = false

[events]
random_numbers_output_enabled = true
unweighting_disabled = true
output_path = training_data_for_flow_model.hdf5
; input_path = random_numbers_from_flow_model.hdf5

[phase_space.vegas]
; disabled = true
alpha = 0.8

[phase_space.optimisation]
n_iter = 15
n_nonzero_min = 80000
n_nonzero_min_growth_factor = 1.00
\end{minted}
\caption{\label{listing:runcard:tt}
	The \Pepper\ runcard used for generating $gg \to t\bar t gggg$ events.
	For generating a different number of final-state gluons, only the \texttt{process} setting needs to be adjusted.}
\end{listing}
\section{Additional figures}
\label{app:additional_plots}
\begin{figure}[tbp]
    \centering
    \includegraphics[width=.48\textwidth]{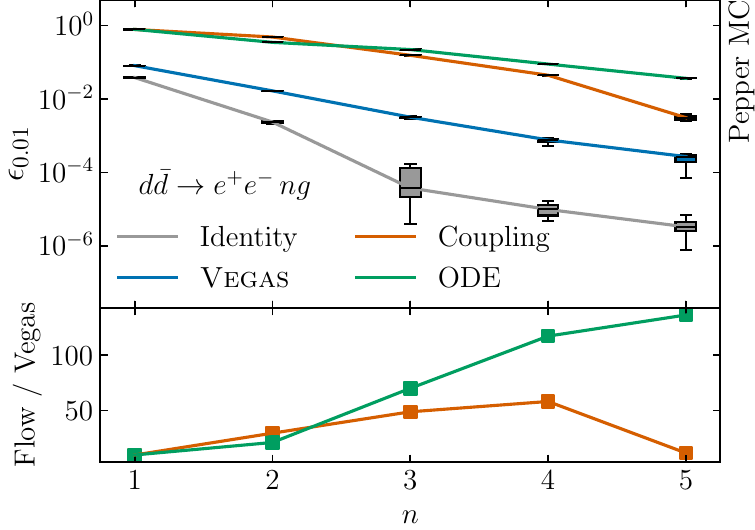}\hfill
    \includegraphics[width=.48\textwidth]{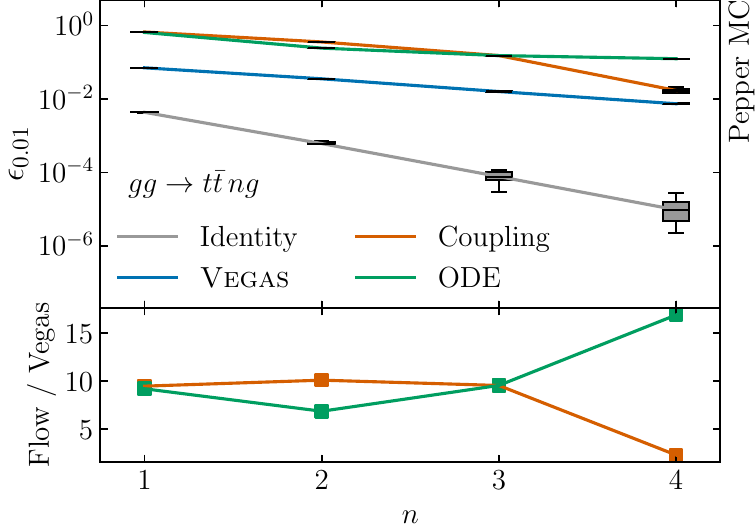}
    \caption{\label{fig:unweff_001}
        Unweighting efficiencies, $\epsilon_{0.01}$ for the \(e^+e^-\) + $n$ gluons
        example (left) and the $t\bar{t}$ + $n$ gluons example (right),
        for various gluon multiplicities $n$.
        The efficiencies are shown for the ``Identity'' mapping,
        and mappings optimised by \Vegas,
        Coupling Flows (``Coupling''),
        and ODE Flows (``ODE'').
    }
\end{figure}

\begin{figure}[tbp]
    \centering
    \includegraphics[width=.48\textwidth]{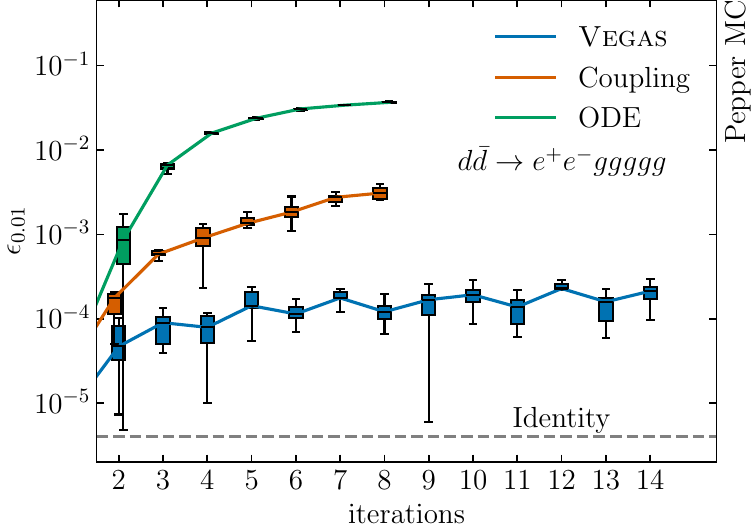}
    \hfill
    \includegraphics[width=.48\textwidth]{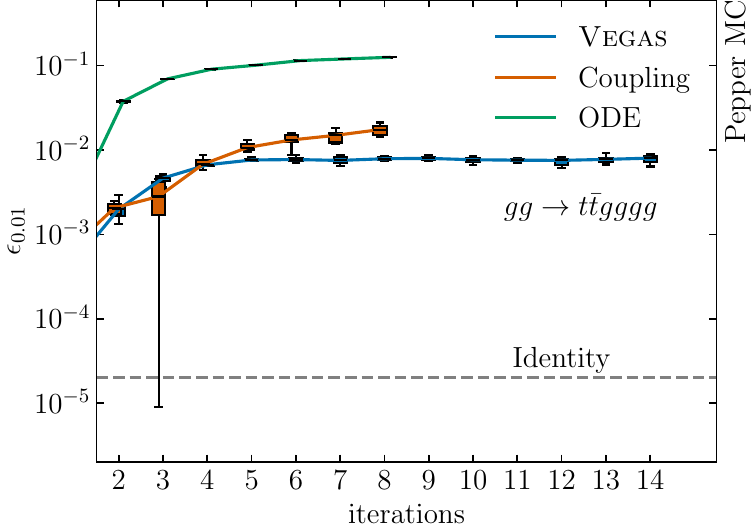}
    \caption{\label{fig:convergence_001}
        Convergence of the effective unweighting efficiency $\epsilon_{\mathrm{0.01}}$
        with training iterations for the highest final-state gluon multiplicities studied: 
        $d\bar d \to e^+e^-+5g$ (left) and 
        $gg \to t\bar t+4g$ (right). Results are shown for the ``Identity'' mapping, Vegas, Coupling Flow (``Coupling''), and ODE Flow (``ODE''). Each curve represents the mean over ten independent evaluations of the unweighting efficiency, the boxes indicate the quartiles of the distribution and the whiskers show $\sfrac{3}{2}$ of the interquartile range.
    }
\end{figure}

In this appendix, we present figures analogous to figure~\ref{fig:unweff} and 
figure~\ref{fig:convergence}, but for $\epsilon_{0.01}$ instead of $\epsilon_{0.001}$. 
figure~\ref{fig:unweff_001} displays the unweighting efficiency as a function of the number
of additional final-state gluons, for both the \(e^+e^-\) + $n$ gluons and $t\bar{t}$ + $n$ gluons
processes, corresponding to figure~\ref{fig:unweff}. Similarly, figure~\ref{fig:convergence_001}
shows the convergence behaviour of the unweighting efficiency for $\epsilon_{0.01}$, 
in correspondence with figure~\ref{fig:convergence}.

\bibliographystyle{JHEP}
\bibliography{literature.bib}

\providecommand{\href}[2]{#2}\begingroup\raggedright\begin{thebibliography}{10}

\bibitem{EuropeanStrategyGroup:2020pow}
{\scshape {European Strategy~Group}} collaboration, \emph{{2020 Update of the
  European Strategy for Particle Physics}}, CERN Council, Geneva (2020),
  \href{https://doi.org/10.17181/ESU2020}{10.17181/ESU2020}.

\bibitem{Narain:2022qud}
M.~Narain et~al., \emph{{The Future of US Particle Physics - The Snowmass 2021
  Energy Frontier Report}},  \href{https://arxiv.org/abs/2211.11084}{{\ttfamily
  2211.11084}}.

\bibitem{HEPSoftwareFoundation:2017ggl}
{\scshape HEP Software Foundation} collaboration, \emph{{A Roadmap for HEP
  Software and Computing R\&D for the 2020s}},
  \href{https://doi.org/10.1007/s41781-018-0018-8}{\emph{Comput. Softw. Big
  Sci.} {\bfseries 3} (2019) 7}
  [\href{https://arxiv.org/abs/1712.06982}{{\ttfamily 1712.06982}}].

\bibitem{HSFPhysicsEventGeneratorWG:2020gxw}
{\scshape HSF Physics Event Generator WG} collaboration, \emph{{Challenges in
  Monte Carlo Event Generator Software for High-Luminosity LHC}},
  \href{https://doi.org/10.1007/s41781-021-00055-1}{\emph{Comput. Softw. Big
  Sci.} {\bfseries 5} (2021) 12}
  [\href{https://arxiv.org/abs/2004.13687}{{\ttfamily 2004.13687}}].

\bibitem{HSFPhysicsEventGeneratorWG:2021xti}
{\scshape HSF Physics Event Generator WG} collaboration, \emph{{HL-LHC
  Computing Review Stage-2, Common Software Projects: Event Generators}},
  \href{https://arxiv.org/abs/2109.14938}{{\ttfamily 2109.14938}}.

\bibitem{ATLAS:2021yza}
{\scshape ATLAS} collaboration, \emph{{Modelling and computational improvements
  to the simulation of single vector-boson plus jet processes for the ATLAS
  experiment}}, \href{https://doi.org/10.1007/JHEP08(2022)089}{\emph{JHEP}
  {\bfseries 08} (2022) 089}
  [\href{https://arxiv.org/abs/2112.09588}{{\ttfamily 2112.09588}}].

\bibitem{Bothmann:2022thx}
E.~Bothmann, A.~Buckley, I.A.~Christidi, C.~G\"utschow, S.~H\"oche, M.~Knobbe
  et~al., \emph{{Accelerating LHC event generation with simplified pilot runs
  and fast PDFs}},
  \href{https://doi.org/10.1140/epjc/s10052-022-11087-1}{\emph{Eur. Phys. J. C}
  {\bfseries 82} (2022) 1128}
  [\href{https://arxiv.org/abs/2209.00843}{{\ttfamily 2209.00843}}].

\bibitem{Kleiss:1994qy}
R.~Kleiss and R.~Pittau, \emph{{Weight optimization in multichannel Monte
  Carlo}}, \href{https://doi.org/10.1016/0010-4655(94)90043-4}{\emph{Comput.
  Phys. Commun.} {\bfseries 83} (1994) 141}
  [\href{https://arxiv.org/abs/hep-ph/9405257}{{\ttfamily hep-ph/9405257}}].

\bibitem{Lepage:1977sw}
G.P.~Lepage, \emph{{A New Algorithm for Adaptive Multidimensional
  Integration}}, \href{https://doi.org/10.1016/0021-9991(78)90004-9}{\emph{J.
  Comput. Phys.} {\bfseries 27} (1978) 192}.

\bibitem{Ohl:1998jn}
T.~Ohl, \emph{{Vegas revisited: Adaptive Monte Carlo integration beyond
  factorization}},
  \href{https://doi.org/10.1016/S0010-4655(99)00209-X}{\emph{Comput. Phys.
  Commun.} {\bfseries 120} (1999) 13}
  [\href{https://arxiv.org/abs/hep-ph/9806432}{{\ttfamily hep-ph/9806432}}].

\bibitem{Lepage:2020tgj}
G.P.~Lepage, \emph{{Adaptive multidimensional integration: VEGAS enhanced}},
  \href{https://doi.org/10.1016/j.jcp.2021.110386}{\emph{J. Comput. Phys.}
  {\bfseries 439} (2021) 110386}
  [\href{https://arxiv.org/abs/2009.05112}{{\ttfamily 2009.05112}}].

\bibitem{Jadach:2002kn}
S.~Jadach, \emph{{Foam: A General purpose cellular Monte Carlo event
  generator}},
  \href{https://doi.org/10.1016/S0010-4655(02)00755-5}{\emph{Comput. Phys.
  Commun.} {\bfseries 152} (2003) 55}
  [\href{https://arxiv.org/abs/physics/0203033}{{\ttfamily physics/0203033}}].

\bibitem{Hahn:2004fe}
T.~Hahn, \emph{{CUBA: A Library for multidimensional numerical integration}},
  \href{https://doi.org/10.1016/j.cpc.2005.01.010}{\emph{Comput. Phys. Commun.}
  {\bfseries 168} (2005) 78}
  [\href{https://arxiv.org/abs/hep-ph/0404043}{{\ttfamily hep-ph/0404043}}].

\bibitem{vanHameren:2007pt}
A.~van Hameren, \emph{{PARNI for importance sampling and density estimation}},
  {\emph{Acta Phys. Polon. B} {\bfseries 40} (2009) 259}
  [\href{https://arxiv.org/abs/0710.2448}{{\ttfamily 0710.2448}}].

\bibitem{Hoche:2019flt}
S.~H\"oche, S.~Prestel and H.~Schulz, \emph{{Simulation of Vector Boson Plus
  Many Jet Final States at the High Luminosity LHC}},
  \href{https://doi.org/10.1103/PhysRevD.100.014024}{\emph{Phys. Rev. D}
  {\bfseries 100} (2019) 014024}
  [\href{https://arxiv.org/abs/1905.05120}{{\ttfamily 1905.05120}}].

\bibitem{Klimek:2018mza}
M.D.~Klimek and M.~Perelstein, \emph{{Neural Network-Based Approach to Phase
  Space Integration}},
  \href{https://doi.org/10.21468/SciPostPhys.9.4.053}{\emph{SciPost Phys.}
  {\bfseries 9} (2020) 053} [\href{https://arxiv.org/abs/1810.11509}{{\ttfamily
  1810.11509}}].

\bibitem{Bothmann:2020ywa}
E.~Bothmann, T.~Jan\ss{}en, M.~Knobbe, T.~Schmale and S.~Schumann,
  \emph{{Exploring phase space with Neural Importance Sampling}},
  \href{https://doi.org/10.21468/SciPostPhys.8.4.069}{\emph{SciPost Phys.}
  {\bfseries 8} (2020) 069} [\href{https://arxiv.org/abs/2001.05478}{{\ttfamily
  2001.05478}}].

\bibitem{Gao:2020zvv}
C.~Gao, S.~H\"oche, J.~Isaacson, C.~Krause and H.~Schulz, \emph{{Event
  Generation with Normalizing Flows}},
  \href{https://doi.org/10.1103/PhysRevD.101.076002}{\emph{Phys. Rev. D}
  {\bfseries 101} (2020) 076002}
  [\href{https://arxiv.org/abs/2001.10028}{{\ttfamily 2001.10028}}].

\bibitem{Heimel:2022wyj}
T.~Heimel, R.~Winterhalder, A.~Butter, J.~Isaacson, C.~Krause, F.~Maltoni
  et~al., \emph{{MadNIS - Neural multi-channel importance sampling}},
  \href{https://doi.org/10.21468/SciPostPhys.15.4.141}{\emph{SciPost Phys.}
  {\bfseries 15} (2023) 141}
  [\href{https://arxiv.org/abs/2212.06172}{{\ttfamily 2212.06172}}].

\bibitem{Verheyen:2022tov}
R.~Verheyen, \emph{{Event Generation and Density Estimation with Surjective
  Normalizing Flows}},
  \href{https://doi.org/10.21468/SciPostPhys.13.3.047}{\emph{SciPost Phys.}
  {\bfseries 13} (2022) 047}
  [\href{https://arxiv.org/abs/2205.01697}{{\ttfamily 2205.01697}}].

\bibitem{Heimel:2023ngj}
T.~Heimel, N.~Huetsch, F.~Maltoni, O.~Mattelaer, T.~Plehn and R.~Winterhalder,
  \emph{{The MadNIS reloaded}},
  \href{https://doi.org/10.21468/SciPostPhys.17.1.023}{\emph{SciPost Phys.}
  {\bfseries 17} (2024) 023}
  [\href{https://arxiv.org/abs/2311.01548}{{\ttfamily 2311.01548}}].

\bibitem{Heimel:2024wph}
T.~Heimel, O.~Mattelaer, T.~Plehn and R.~Winterhalder, \emph{{Differentiable
  MadNIS-Lite}},
  \href{https://doi.org/10.21468/SciPostPhys.18.1.017}{\emph{SciPost Phys.}
  {\bfseries 18} (2025) 017}
  [\href{https://arxiv.org/abs/2408.01486}{{\ttfamily 2408.01486}}].

\bibitem{Butter:2022rso}
S.~Badger et~al., \emph{{Machine learning and LHC event generation}},
  \href{https://doi.org/10.21468/SciPostPhys.14.4.079}{\emph{SciPost Phys.}
  {\bfseries 14} (2023) 079}
  [\href{https://arxiv.org/abs/2203.07460}{{\ttfamily 2203.07460}}].

\bibitem{Chen:2018}
R.T.Q.~Chen, Y.~Rubanova, J.~Bettencourt and D.~Duvenaud, \emph{Neural ordinary
  differential equations},  in \emph{Proceedings of the 32nd International
  Conference on Neural Information Processing Systems}, NeurIPS'18, (Red Hook,
  NY, USA), p.~6572–6583, Curran Associates Inc., 2018
  [\href{https://arxiv.org/abs/1806.07366}{{\ttfamily 1806.07366}}].

\bibitem{Lipman:2023}
Y.~Lipman, R.T.Q.~Chen, H.~Ben-Hamu, M.~Nickel and M.~Le, \emph{Flow matching
  for generative modeling},  \href{https://arxiv.org/abs/2210.02747}{{\ttfamily
  2210.02747}}.

\bibitem{Albergo:2023building}
M.S.~Albergo and E.~Vanden-Eijnden, \emph{Building normalizing flows with
  stochastic interpolants},  in \emph{The Eleventh International Conference on
  Learning Representations}, 2023
  [\href{https://arxiv.org/abs/2209.15571}{{\ttfamily 2209.15571}}].

\bibitem{Albergo:2023stochastic}
M.S.~Albergo, N.M.~Boffi and E.~Vanden-Eijnden, \emph{Stochastic interpolants:
  A unifying framework for flows and diffusions},
  \href{https://arxiv.org/abs/2303.08797}{{\ttfamily 2303.08797}}.

\bibitem{Liu:2022}
X.~Liu, C.~Gong and Q.~Liu, \emph{Flow straight and fast: Learning to generate
  and transfer data with rectified flow},
  \href{https://arxiv.org/abs/2209.03003}{{\ttfamily 2209.03003}}.

\bibitem{Butter:2023fov}
A.~Butter, N.~Huetsch, S.~Palacios~Schweitzer, T.~Plehn, P.~Sorrenson and
  J.~Spinner, \emph{{Jet Diffusion versus JetGPT -- Modern Networks for the
  LHC}},  \href{https://arxiv.org/abs/2305.10475}{{\ttfamily 2305.10475}}.

\bibitem{Favaro:2025psi}
L.~Favaro, R.~Kogler, A.~Paasch, S.~Palacios~Schweitzer, T.~Plehn and
  D.~Schwarz, \emph{{How to Unfold Top Decays}},
  \href{https://arxiv.org/abs/2501.12363}{{\ttfamily 2501.12363}}.

\bibitem{Bothmann:2023gew}
E.~Bothmann, T.~Childers, W.~Giele, S.~H\"oche, J.~Isaacson and M.~Knobbe,
  \emph{{A portable parton-level event generator for the high-luminosity LHC}},
  \href{https://doi.org/10.21468/SciPostPhys.17.3.081}{\emph{SciPost Phys.}
  {\bfseries 17} (2024) 081}
  [\href{https://arxiv.org/abs/2311.06198}{{\ttfamily 2311.06198}}].

\bibitem{Bothmann:2023siu}
E.~Bothmann, T.~Childers, W.~Giele, F.~Herren, S.~Hoeche, J.~Isaacson et~al.,
  \emph{{Efficient phase-space generation for hadron collider event
  simulation}},
  \href{https://doi.org/10.21468/SciPostPhys.15.4.169}{\emph{SciPost Phys.}
  {\bfseries 15} (2023) 169}
  [\href{https://arxiv.org/abs/2302.10449}{{\ttfamily 2302.10449}}].

\bibitem{Dinh:2014}
L.~Dinh, D.~Krueger and Y.~Bengio, \emph{{NICE:} non-linear independent
  components estimation},  in \emph{3rd International Conference on Learning
  Representations, {ICLR} 2015, San Diego, CA, USA, May 7-9, 2015, Workshop
  Track Proceedings}, Y.~Bengio and Y.~LeCun, eds., 2015,
  \href{http://arxiv.org/abs/1410.8516}{http://arxiv.org/abs/1410.8516}
  [\href{https://arxiv.org/abs/1410.8516}{{\ttfamily 1410.8516}}].

\bibitem{Mueller:2019}
T.~M\"{u}ller, B.~Mcwilliams, F.~Rousselle, M.~Gross and J.~Nov\'{a}k,
  \emph{Neural importance sampling},
  \href{https://doi.org/10.1145/3341156}{\emph{ACM Trans. Graph.} {\bfseries
  38} (2019) } [\href{https://arxiv.org/abs/1808.03856}{{\ttfamily
  1808.03856}}].

\bibitem{Durkan:2019}
C.~Durkan, A.~Bekasov, I.~Murray and G.~Papamakarios, \emph{Neural spline
  flows},  in \emph{Advances in Neural Information Processing Systems},
  H.~Wallach, H.~Larochelle, A.~Beygelzimer, F.~d\textquotesingle
  Alch\'{e}-Buc, E.~Fox and R.~Garnett, eds., vol.~32, Curran Associates, Inc.,
  2019 [\href{https://arxiv.org/abs/1906.04032}{{\ttfamily 1906.04032}}].

\bibitem{1961_Neumann_John}
J.~von Neumann, \emph{{John von Neumann Collected Works}},  vol.~{5: Design of
  Computers, Theory of Automata and Numerical Analysis}, (Oxford, England),
  pp.~768--770, Pergamon Press (1961).

\bibitem{kish1965survey}
L.~Kish, \emph{Survey Sampling}, Wiley (1965).

\bibitem{Tabak:2010}
E.~Tabak and E.~Vanden-Eijnden, \emph{Density estimation by dual ascent of the
  log-likelihood},
  \href{https://doi.org/10.4310/CMS.2010.v8.n1.a11}{\emph{Communications in
  Mathematical Sciences - COMMUN MATH SCI} {\bfseries 8} (2010) }.

\bibitem{Tabak:2013}
E.G.~Tabak and C.V.~Turner, \emph{A family of nonparametric density estimation
  algorithms},
  \href{https://doi.org/https://doi.org/10.1002/cpa.21423}{\emph{Communications
  on Pure and Applied Mathematics} {\bfseries 66} (2013) 145}.

\bibitem{Stienen:2021gns}
B.~Stienen and R.~Verheyen, \emph{{Phase space sampling and inference from
  weighted events with autoregressive flows}},
  \href{https://doi.org/10.21468/SciPostPhys.10.2.038}{\emph{SciPost Phys.}
  {\bfseries 10} (2021) 038}
  [\href{https://arxiv.org/abs/2011.13445}{{\ttfamily 2011.13445}}].

\bibitem{Deutschmann:2024lml}
N.~Deutschmann and N.~G\"otz, \emph{{Accelerating HEP simulations with Neural
  Importance Sampling}},
  \href{https://doi.org/10.1007/JHEP03(2024)083}{\emph{JHEP} {\bfseries 03}
  (2024) 083} [\href{https://arxiv.org/abs/2401.09069}{{\ttfamily
  2401.09069}}].

\bibitem{Finlay:2020}
C.~Finlay, J.-H.~Jacobsen, L.~Nurbekyan and A.M.~Oberman, \emph{How to train
  your neural ode: the world of jacobian and kinetic regularization},  in
  \emph{Proceedings of the 37th International Conference on Machine Learning},
  ICML'20, JMLR.org, 2020 [\href{https://arxiv.org/abs/2002.02798}{{\ttfamily
  2002.02798}}].

\bibitem{Onken:2021}
D.~Onken, S.W.~Fung, X.~Li and L.~Ruthotto, \emph{{OT-Flow}: Fast and accurate
  continuous normalizing flows via optimal transport},  in \emph{AAAI
  Conference on Artificial Intelligence}, vol.~35, pp.~9223--9232, May, 2021,
  \href{https://ojs.aaai.org/index.php/AAAI/article/view/17113}{https://ojs.aaai.org/index.php/AAAI/article/view/17113}
  [\href{https://arxiv.org/abs/2006.00104}{{\ttfamily 2006.00104}}].

\bibitem{Tong:2024}
A.~Tong, K.~FATRAS, N.~Malkin, G.~Huguet, Y.~Zhang, J.~Rector-Brooks et~al.,
  \emph{Improving and generalizing flow-based generative models with minibatch
  optimal transport}, {\emph{Transactions on Machine Learning Research} (2024)
  } [\href{https://arxiv.org/abs/2302.00482}{{\ttfamily 2302.00482}}].

\bibitem{Pooladian:2023}
A.-A.~Pooladian, H.~Ben-Hamu, C.~Domingo-Enrich, B.~Amos, Y.~Lipman and
  R.T.Q.~Chen, \emph{Multisample flow matching: straightening flows with
  minibatch couplings},  in \emph{Proceedings of the 40th International
  Conference on Machine Learning}, ICML'23, JMLR.org, 2023
  [\href{https://arxiv.org/abs/2304.14772}{{\ttfamily 2304.14772}}].

\bibitem{LaCagnina:2024wcc}
S.~La~Cagnina, C.~Grunwald, T.~Jan\ss{}en, K.~Kr\"oninger and S.~Schumann,
  \emph{{Phase space sampling with Markov Chain Monte Carlo methods}},  12,
  2024 [\href{https://arxiv.org/abs/2412.12963}{{\ttfamily 2412.12963}}].

\bibitem{Janssen:2025zke}
T.~Jan\ss{}en, R.~Poncelet and S.~Schumann, \emph{{Sampling NNLO QCD phase
  space with normalizing flows}},
  \href{https://arxiv.org/abs/2505.13608}{{\ttfamily 2505.13608}}.

\bibitem{Bothmann:2021nch}
E.~Bothmann, W.~Giele, S.~Hoeche, J.~Isaacson and M.~Knobbe, \emph{{Many-gluon
  tree amplitudes on modern GPUs: A case study for novel event generators}},
  \href{https://doi.org/10.21468/SciPostPhysCodeb.3}{\emph{SciPost Phys.
  Codeb.} {\bfseries 2022} (2022) 3}
  [\href{https://arxiv.org/abs/2106.06507}{{\ttfamily 2106.06507}}].

\bibitem{Bothmann:2022itv}
E.~Bothmann, J.~Isaacson, M.~Knobbe, S.~H\"oche and W.~Giele, \emph{{QCD tree
  amplitudes on modern GPUs: A case study for novel event generators}},
  \href{https://doi.org/10.22323/1.414.0222}{\emph{PoS} {\bfseries ICHEP2022}
  (2022) 222}.

\bibitem{Gleisberg:2008fv}
T.~Gleisberg and S.~Hoeche, \emph{{Comix, a new matrix element generator}},
  \href{https://doi.org/10.1088/1126-6708/2008/12/039}{\emph{JHEP} {\bfseries
  12} (2008) 039} [\href{https://arxiv.org/abs/0808.3674}{{\ttfamily
  0808.3674}}].

\bibitem{Gleisberg:2008ta}
T.~Gleisberg, S.~H{\"o}che, F.~Krauss, M.~Sch{\"o}nherr, S.~Schumann,
  F.~Siegert et~al., \emph{{Event generation with SHERPA 1.1}},
  \href{https://doi.org/10.1088/1126-6708/2009/02/007}{\emph{JHEP} {\bfseries
  02} (2009) 007} [\href{https://arxiv.org/abs/0811.4622}{{\ttfamily
  0811.4622}}].

\bibitem{Sherpa:2019gpd}
{\scshape Sherpa} collaboration, \emph{{Event Generation with Sherpa 2.2}},
  \href{https://doi.org/10.21468/SciPostPhys.7.3.034}{\emph{SciPost Phys.}
  {\bfseries 7} (2019) 034} [\href{https://arxiv.org/abs/1905.09127}{{\ttfamily
  1905.09127}}].

\bibitem{Sherpa:2024mfk}
{\scshape Sherpa} collaboration, \emph{{Event generation with Sherpa 3}},
  \href{https://doi.org/10.1007/JHEP12(2024)156}{\emph{JHEP} {\bfseries 12}
  (2024) 156} [\href{https://arxiv.org/abs/2410.22148}{{\ttfamily
  2410.22148}}].

\bibitem{Melia:2013bta}
T.~Melia, \emph{{Dyck words and multiquark primitive amplitudes}},
  \href{https://doi.org/10.1103/PhysRevD.88.014020}{\emph{Phys. Rev. D}
  {\bfseries 88} (2013) 014020}
  [\href{https://arxiv.org/abs/1304.7809}{{\ttfamily 1304.7809}}].

\bibitem{Melia:2013xok}
T.~Melia, \emph{{Dyck words and multi-quark amplitudes}},
  \href{https://doi.org/10.22323/1.197.0031}{\emph{PoS} {\bfseries RADCOR2013}
  (2013) 031}.

\bibitem{Johansson:2015oia}
H.~Johansson and A.~Ochirov, \emph{{Color-Kinematics Duality for QCD
  Amplitudes}}, \href{https://doi.org/10.1007/JHEP01(2016)170}{\emph{JHEP}
  {\bfseries 01} (2016) 170}
  [\href{https://arxiv.org/abs/1507.00332}{{\ttfamily 1507.00332}}].

\bibitem{hdf5}
{The HDF Group}, \emph{{Hierarchical Data Format, version 5}},
  \href{https://www.hdfgroup.org/HDF5/}{https://www.hdfgroup.org/HDF5/}.

\bibitem{highfive}
\emph{{HighFive - HDF5 header-only C++ Library}},
  \href{https://bluebrain.github.io/HighFive/}{https://bluebrain.github.io/HighFive/}.

\bibitem{Hoeche:2019rti}
S.~Höche, S.~Prestel and H.~Schulz, \emph{{Simulation of Vector Boson Plus
  Many Jet Final States at the High Luminosity LHC}},
  \href{https://doi.org/10.1103/PhysRevD.100.014024}{\emph{Phys. Rev.}
  {\bfseries D100} (2019) 014024}
  [\href{https://arxiv.org/abs/1905.05120}{{\ttfamily 1905.05120}}].

\bibitem{Bothmann:2023ozs}
E.~Bothmann, T.~Childers, C.~G\"utschow, S.~H\"oche, P.~Hovland, J.~Isaacson
  et~al., \emph{{Efficient precision simulation of processes with many-jet
  final states at the LHC}},
  \href{https://doi.org/10.1103/PhysRevD.109.014013}{\emph{Phys. Rev. D}
  {\bfseries 109} (2024) 014013}
  [\href{https://arxiv.org/abs/2309.13154}{{\ttfamily 2309.13154}}].

\bibitem{NNPDF:2014otw}
{\scshape NNPDF} collaboration, \emph{{Parton distributions for the LHC Run
  II}}, \href{https://doi.org/10.1007/JHEP04(2015)040}{\emph{JHEP} {\bfseries
  04} (2015) 040} [\href{https://arxiv.org/abs/1410.8849}{{\ttfamily
  1410.8849}}].

\bibitem{Buckley:2014ana}
A.~Buckley, J.~Ferrando, S.~Lloyd, K.~Nordstr\"om, B.~Page, M.~R\"ufenacht
  et~al., \emph{{LHAPDF6: parton density access in the LHC precision era}},
  \href{https://doi.org/10.1140/epjc/s10052-015-3318-8}{\emph{Eur. Phys. J. C}
  {\bfseries 75} (2015) 132} [\href{https://arxiv.org/abs/1412.7420}{{\ttfamily
  1412.7420}}].

\bibitem{Bern:2013gka}
Z.~Bern, L.J.~Dixon, F.~Febres~Cordero, S.~H\"oche, H.~Ita, D.A.~Kosower
  et~al., \emph{{Next-to-Leading Order $W + 5$-Jet Production at the LHC}},
  \href{https://doi.org/10.1103/PhysRevD.88.014025}{\emph{Phys. Rev. D}
  {\bfseries 88} (2013) 014025}
  [\href{https://arxiv.org/abs/1304.1253}{{\ttfamily 1304.1253}}].

\bibitem{Akhound:2024}
T.~Akhound-Sadegh, J.~Rector-Brooks, A.~Joey~Bose, S.~Mittal, P.~Lemos,
  C.-H.~Liu et~al., \emph{Iterated denoising energy matching for sampling from
  boltzmann densities},  in \emph{Proceedings of the 41st International
  Conference on Machine Learning}, ICML'24, JMLR.org, 2024
  [\href{https://arxiv.org/abs/2402.06121}{{\ttfamily 2402.06121}}].

\bibitem{Gao:2020vdv}
C.~Gao, J.~Isaacson and C.~Krause, \emph{{i-flow: High-dimensional Integration
  and Sampling with Normalizing Flows}},
  \href{https://doi.org/10.1088/2632-2153/abab62}{\emph{Mach. Learn. Sci.
  Tech.} {\bfseries 1} (2020) 045023}
  [\href{https://arxiv.org/abs/2001.05486}{{\ttfamily 2001.05486}}].

\bibitem{nflows:2020}
C.~Durkan, A.~Bekasov, I.~Murray and G.~Papamakarios, \emph{{nflows}:
  normalizing flows in {PyTorch}}, Zenodo (2020),
  \href{https://doi.org/10.5281/zenodo.4296287}{10.5281/zenodo.4296287}.

\bibitem{adamw:2019}
I.~Loshchilov and F.~Hutter, \emph{Decoupled weight decay regularization},  in
  \emph{7th International Conference on Learning Representations, {ICLR} 2019,
  New Orleans, LA, USA, May 6-9, 2019}, OpenReview.net, 2019
  [\href{https://arxiv.org/abs/1711.05101}{{\ttfamily 1711.05101}}].

\bibitem{Wildberger:2023}
J.~Wildberger, M.~Dax, S.~Buchholz, S.R.~Green, J.H.~Macke and
  B.~Sch\"{o}lkopf, \emph{Flow matching for scalable simulation-based
  inference},  in \emph{Proceedings of the 37th International Conference on
  Neural Information Processing Systems}, NeurIPS '23, (Red Hook, NY, USA),
  Curran Associates Inc., 2023
  [\href{https://arxiv.org/abs/2305.17161}{{\ttfamily 2305.17161}}].

\bibitem{Tancik:2020}
M.~Tancik, P.P.~Srinivasan, B.~Mildenhall, S.~Fridovich-Keil, N.~Raghavan,
  U.~Singhal et~al., \emph{Fourier features let networks learn high frequency
  functions in low dimensional domains},  in \emph{Proceedings of the 34th
  International Conference on Neural Information Processing Systems}, NeurIPS
  '20, (Red Hook, NY, USA), Curran Associates Inc., 2020
  [\href{https://arxiv.org/abs/2006.10739}{{\ttfamily 2006.10739}}].

\bibitem{DORMAND198019}
J.~Dormand and P.~Prince, \emph{A family of embedded runge-kutta formulae},
  \href{https://doi.org/https://doi.org/10.1016/0771-050X(80)90013-3}{\emph{Journal
  of Computational and Applied Mathematics} {\bfseries 6} (1980) 19}.

\bibitem{politorchdyn}
M.~Poli, S.~Massaroli, A.~Yamashita, H.~Asama, J.~Park and S.~Ermon,
  \emph{Torchdyn: Implicit models and neural numerical methods in pytorch},
  \href{https://github.com/DiffEqML/torchdyn}{https://github.com/DiffEqML/torchdyn}.

\bibitem{Rehman:2025}
D.~Rehman, O.~Davis, J.~Lu, J.~Tang, M.~Bronstein, Y.~Bengio et~al.,
  \emph{Fort: Forward-only regression training of normalizing flows},
  \href{https://arxiv.org/abs/2506.01158}{{\ttfamily 2506.01158}}.

\end{thebibliography}\endgroup

\end{document}